# Voltage-Current-Emission Framework of Time-Resolved Electroluminescence

Hui Bao[1], Longjia Wu[2], Yanan Guo[3], Jianchang Yan[4], Ying Wang[5], Haizheng Zhong[6]*

[1]MIIT Key Laboratory for Low-Dimensional Quantum Structure and Devices, School of Optics and Photonics, Beijing Institute of Technology, Beijing 100081, China.

[2]New Display Technology Department, TCL Corporate Research, Guangzhou, Guangdong, 510000, China.

[3]Research and Development Center for Wide Bandgap Semiconductors, Institute of Semiconductors, Chinese Academy of Sciences, Beijing 100083, China.

[4]School of Integrated Circuit Science and Engineering, Beihang University, Haidian District, Beijing 100191, China.

[5]Key Laboratory of Photochemical Conversion and Optoelectronic Materials and CityU−CAS Joint Laboratory of Functional Materials and Devices, Technical Institute of Physics and Chemistry, Chinese Academy of Sciences, Beijing, China.

[6]MIIT Key Laboratory for Low-Dimensional Quantum Structure and Devices, School of Materials Science & Engineering, Beijing Institute of Technology, Beijing 100081, China.

*Corresponding Author's E-mail: hzzhong@bit.edu.cn

**Abstract:** Time-resolved electroluminescence has been a central methodology in the development of LED devices. However, it still lacks quantitative theoretical descriptions of time-resolved electroluminescence. By combining the equivalent circuit models of semiconductor devices and the carrier dynamics of electroluminescence processes, we here developed a Voltage-Current-Emission framework to illustrate time-resolved electroluminescence from the pulse voltage as input, to injection current and finally generate electroluminescence as output. The features of different LED devices can be quantitatively interpreted, confirming the universal applicability across LEDs with different materials and device structures. Importantly, the curves of time-resolved electroluminescence can be simulated to quantitatively describe the dynamics of current and carriers of LED operation. In all, the Voltage-Current-Emission framework offers theoretical guidance for achieving advanced LED devices for AI technology.

## Main

Light emitting diodes (LED) have been central devices to revolute lighting and display technology[1-3], providing products of GaN-LED[4,5], organic light emitting diode (OLED)[6], and demos of quantum dots light emitting diode (QLED)[7]. With the revolution of AI technology, there are emerging demands to develop advanced LED devices with high resolution, high brightness, and/or high response speed for AR display[8-11], optical communication[12], and optical computer[13]. Time-resolved electroluminescence (TREL) measures the electroluminescence (EL) signals over time under an applied pulse voltage, while time-resolved current (TRC) can record the evolution of current with time prolonging[14-19]. Previous analysis of TREL curves have tried to illustrate the carrier dynamics of LED operation from the microscopic viewpoint including recombination, transport and injection[17,20-25]. Equivalent circuit models have been widely used in the analysis of semiconductor devices but have been neglected in TREL interpretations[25-29]. Consequently, it still lacks a comprehensive picture to quantitatively describe the TREL curves.

Typically, the TREL curves of LED devices usually contain four stages including delay, rising, balance, and decay[16,25]. Delay stage shows immediate current flow under voltage application while no detectable luminescence is observed. During rising stage, the time-resolved current approaches to a balance state, meanwhile the EL signals rises and approach to maximum value with time prolonging. Balance stage is similar with operation state under constant voltage. Decay stage reveals the features of luminescence decay after pulse voltage. It is worth noting that there are also conflicts in describing TREL stages. For example, the rising edge has been attributed to the rise of electron concentration[23], hole concentration[24], or both[25]. The delay stage was correlated to the transport time of low-mobility carriers[30,31,32], or the charging time of device capacitance[25], or their combination[33,22]. These limitations have led to many misunderstandings in the interpretation of TREL.

We here discover the operational principle of "voltage drives current, current drives emission" in LED devices. In response to an applied pulse voltage, LED devices immediately generate current, and then the injected current leads to the formation of excitons with following emission. Based on this principle, the TREL dynamics can be correlated with time-resolved current using equivalent circuit models. To verify this point, we varied external resistors (variable control 1), low-level voltages (variable control 2), and layer thicknesses of hole transport layer, emissive layer and electron transport layer (variable control 3,4,5) to experimentally connect the equivalent circuit model and carrier dynamics of LEDs using QLED as a platform (Figure 1a), and then the conclusions are verified in GaN-LED and OLED.

Figure 1b shows the typical TREL curves of QLED under different external resistors. To ensure that the voltage drop across the LED device remained constant, the applied pulse voltage was adjusted accordingly when the external resistor was varied. With external resistors increasing from 100 Ω to 1100 Ω, delay time ($T_d$) increases from 0.85 μs to 1.70 μs and decay time ($T_c$) increases from 0.44 μs to 0.81 μs. As shown in the inset of Fig. 1a, $V_L$ represents the voltage level of the pulse signal at its low state and $V_H$ represents the voltage level of the pulse signal at its high state. Figure 1c shows the relationship between delay time ($T_d$) and the low-level voltage ($V_L$) of the pulse signal, with its high-level voltage ($V_H$) maintained constant. Delay time ($T_d$) decreases from 1.0 μs to near zero with the $V_L$ increasing from 0V to 2.0V. As shown in Fig. 1d, when the $V_L$ gets close to the device's turn-on voltage, $T_d$ almost approaches to zero. Figure 1e shows the delay stage of TREL curves for QLED with QD layers of different thickness. When the QD layer becomes thicker, delay time ($T_d$) decreases from 0.73 μs to 0.51 μs. Similarly, $T_d$ decreases from 0.91 μs to 0.49 μs with the thicknesses of hole transport layer (HTL) increasing (Fig. S1a). Figures 1f and S1b show the decay time for different electron transport layer (ETL) and QD thicknesses. The decay time decreases with increasing ETL or QD thickness. These trends were also observed in OLED and GaN-LED devices (Figs. S2, S3). In all, the delay and decay time are not only related with material properties, but also correlated with external circuit, low-level voltage and device structure. Based on these results, both of the carrier dynamics (microscopic) and equivalent circuit (macroscopic) can vary the TREL behaviors.

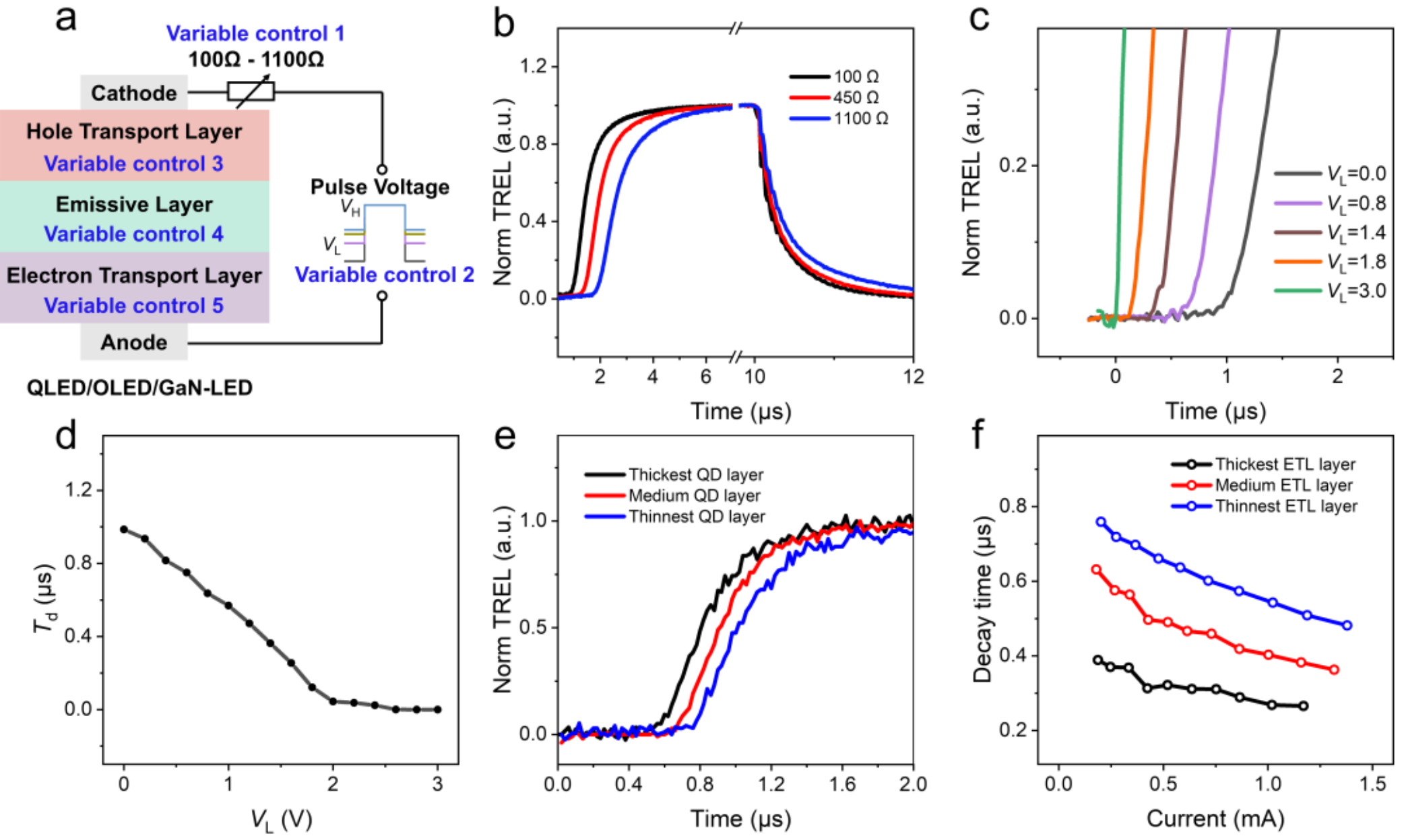

**Figure 1. a)** Schematic diagram of the TREL test circuit and variable control. **b)** TREL response for QLED with external resistances of 100 Ω, 450 Ω, and 1100 Ω. **c)** Delay phase characteristics of TREL curves under different $V_L$ voltages. **d)** Variation of TREL curve delay time with $V_L$ voltage in QLEDs. **e)** TREL curves of QLED devices with different quantum dot layer thicknesses. **f)** Decay time of TREL curves versus equilibrium current for QLEDs with different ETL thicknesses.

To gain a comprehensive physical picture, we propose physical framework of Voltage-Current-Emission (VCE) to describe the evolution from voltage input to current injection and ultimately light emission. As shown in Fig. 2, it considers time-resolved injection current by introducing a diode into RC equivalent circuit model, which connects the equivalent circuit model and carrier dynamics.

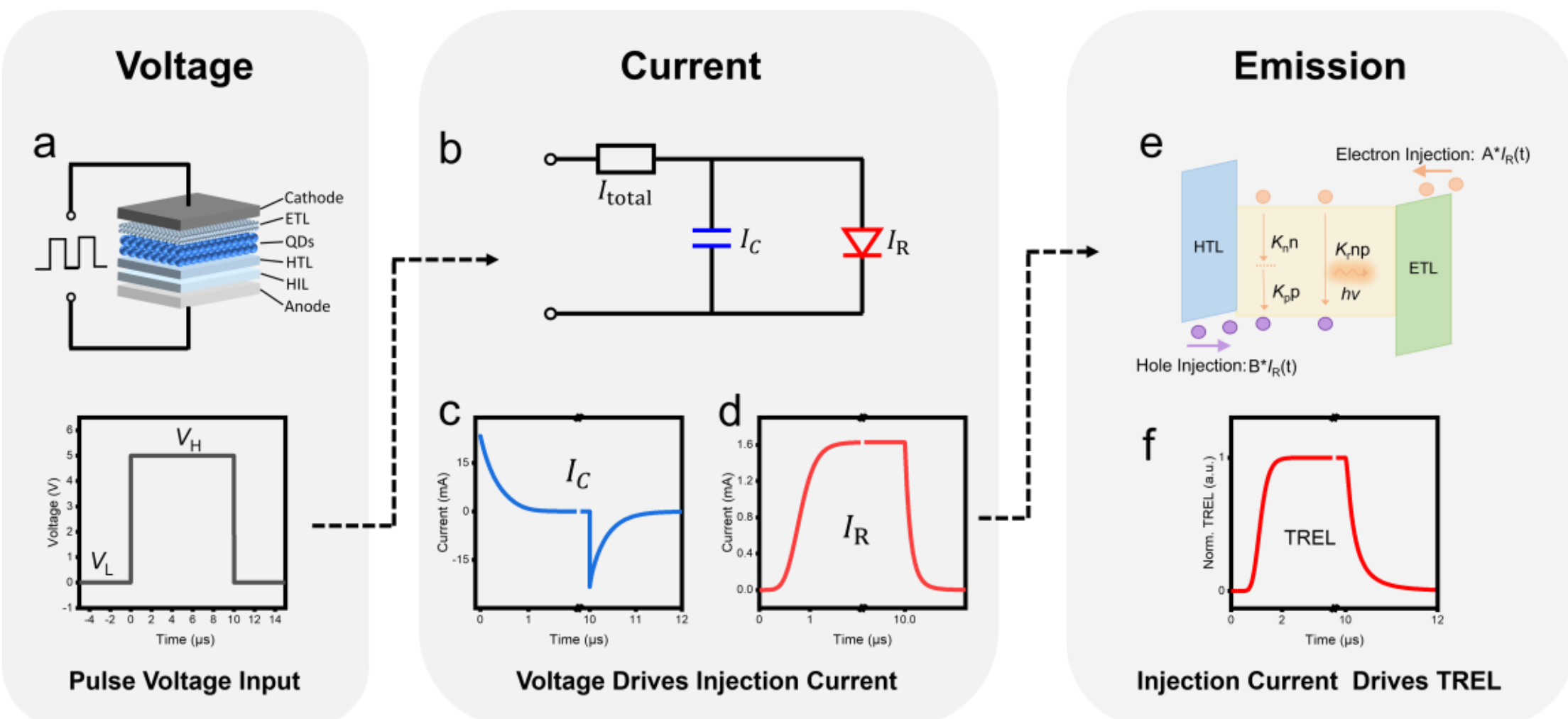

**Figure 2. Schematic of the Voltage-Current-Emission (VCE) framework. a**) Device structure and pulse voltage input. **b**) Diode-based RC equivalent circuit. **c**) Simulated capacitor current ($I_C$) as a function of time. **d**) Simulated diode current ($I_R$) as a function of time. **e**) Schematical figure of carrier dynamics in the emissive layer. **f**) Simulated diode current and corresponding TREL curve.

For a LED device, the injection current exhibits an exponential relationship with voltage, which can be described using Shockley equation[29,34-36]. As shown in Fig. 2b, we used a diode-based RC equivalent circuit model for transient current analysis. Compared with the linear RC parallel model, the key improvement of diode-based RC model lies in its accurate characterization of the nonlinear nature of LED. The dynamics of the improved equivalent circuit system can be described by the following set of equations.

$$\frac{1}{C}\int_0^t I_C dt = \frac{Q}{C} = V_R \tag{1}$$

$$I_R + I_C = I_{total} \tag{2}$$

$$V_R + R_s * I_{total} = V \tag{3}$$

$$I_R = I_{sat}\,(e^{\frac{qV_R}{nkT}} - 1) \tag{4}$$

Here, $I_R$, $I_C$, $I_{total}$, $V$, $Q$, and $V_R$ represent the PN junction current, capacitor current, total current, applied voltage, accumulated charge on the capacitor, and voltage across parallel components respectively, with $I_{sat}$ and $n$ denoting the saturation current and ideality factor. Since these coupled differential equations lack an analytical solution, we use numerical simulation to obtain the transient response of this equivalent circuit. Figure 2c and Figure 2d show the simulated time-dependent curves of $I_R$ and $I_C$ ($I_{sat} = 5.52\, x 10^{-6} mA$, $n = 20$, $C = 1.2nf$, $R_s = 300\Omega$).

Under voltage pulse application, $I_C$ instantly reaches peak values and then decays to balance values due to the capacitor charging process, while $I_R$ exhibits a delayed process (defined as current delay time) and then increases to a balance value. After pulse termination, $I_C$ immediately reverses to peak negative values before decaying to zero due to the capacitor discharging process, whereas $I_R$ decays from its original value to zero (defined as current decay time). Owing to the PN junction features of LED device, PN junction current ($I_R$) serves as the injection current for both electrons and holes in subsequent TREL simulations.

TREL curves represent the time evolution of carrier concentration in the emissive layer, whose quantitative description is derived from carrier continuity equations, as shown in Fig. 2e. TREL measurements only detect temporal dynamics, allowing us to ignore spatial terms and thus simplify the continuity equations to following form.

$$\frac{dn(t)}{dt} = I_{\mathrm{n}}(\mathrm{t}) - k_r n(t)p(t) - k_n n(\mathrm{t}) \tag{5a}$$

$$\frac{dp(t)}{dt} = I_{\mathrm{p}}(\mathrm{t}) - k_r n(t)p(t) - k_p p(\mathrm{t}) \tag{5b}$$

$$TREL_{norm} = \frac{n(t)p(t)}{(np)_{max}} \tag{6}$$

Here, $n(t)$ and $p(t)$ represent the total carrier counts in the quantum dot layer, converted from carrier concentrations through volume of QD layer. $I_{\mathrm{n}}(\mathrm{t})$ and $I_{\mathrm{p}}(\mathrm{t})$ represent electron and hole injection rates (units: carriers/μs), defined as $I_{\mathrm{n}}(\mathrm{t}) = A * I_{\mathrm{R}}(\mathrm{t})/q$ and $I_{\mathrm{p}}(\mathrm{t}) = B * I_{\mathrm{R}}(\mathrm{t})/q$, where A and B (0<A, B<1) represent the ratios of electron and hole injection currents relative to the total diode current $I_{\mathrm{R}}(\mathrm{t})$. The second order recombination term $k_r n(t)p(t)$ describes bimolecular recombination kinetics, where $k_r$ is the bimolecular recombination coefficient. First order loss terms $k_n n(\mathrm{t})$, $k_p p(\mathrm{t})$ include deep-trap-assisted non-radiative recombination and carrier leakage effects, both exhibiting first order kinetics. It is noted that shallow defect-induced non-radiative recombination is also a bimolecular process. Thus, bimolecular recombination coefficient cannot be considered as the radiative recombination coefficient. Figure 2f presents the simulated diode current and its corresponding TREL simulation results. When the diode current is substituted into equations (5) and (6) for simulation, a time delay between diode current and TREL can be noticed. This observation can be explained to the information masking effects of normalization at the initial injection stage.

Based on VCE framework, we qualitatively illustrate the evolution of current and emission under different stages. In delay stage, capacitor charging occurs without diode current, while the diode voltages gradually increasing to turn-on voltage. In rising stage, the capacitor continues to be charged until it reaches saturated value, while the diode's current increases along with EL rising. In decay stage, the fully charged capacitor discharges, while the diode's current decays along with EL decay. In the following, we tried to explain the experimental results through numerical simulation and fitting.

Figures. 3a and 3b simulated the influences of capacitance and external resistance on the diode current. Either the increase of capacitance or external resistance significantly prolong the delay time of the diode injection current, which also increase the decay time of diode current. Figs. 3c and 3d show the simulated TREL curves. With the capacitance or external resistance increasing, both of the delay and decay time of TREL become longer. Figure 3e presents simulated TREL curves under varying $V_{\mathrm{L}}$. The increase of $V_{\mathrm{L}}$ reduces the delay time. Moreover, the delay time starts to disappear when the $V_{\mathrm{L}}$ approaches the turn-on voltage (Fig. 3f). The good agreements between simulations (Fig. 3) and experiments (Fig. 1) demonstrate that the experimental TRC and TREL results can be well described using the VCE framework.

Specifically, the delay time of TREL can be explained to the current delay of diode with turning-on and capacitor charging, while the decay time of TREL is correlated with the current decay of diode with capacitor discharging. Before the turn-on voltage, no current can be observed in the delay stage of diode current due to capacitor charging, which leads to the delay of TREL. Therefore, in the aforementioned experimental results, when reducing the thickness of the transport layer or increase the external series resistance, the increased charging time of the capacitance leads to a longer delay in the diode current, consequently increasing the TREL delay time. Similarly, as $V_{\mathrm{L}}$ increases, the initial voltage division across the parallel capacitor increases, thus reduces the time required for the capacitor to charge to the turn-on voltage. Consequently, the delay in diode current is shortened, leading to a decrease in the TREL delay time (Fig. 1c, Fig.3e). Notably, when $V_{\mathrm{L}}$ exceeds the turn-on voltage, no charging time is required, and delay time drops to zero (Fig. 1d, Fig.3f). In comparison, the decay stage of TREL is correlated with diode current decaying due to capacitor discharging.

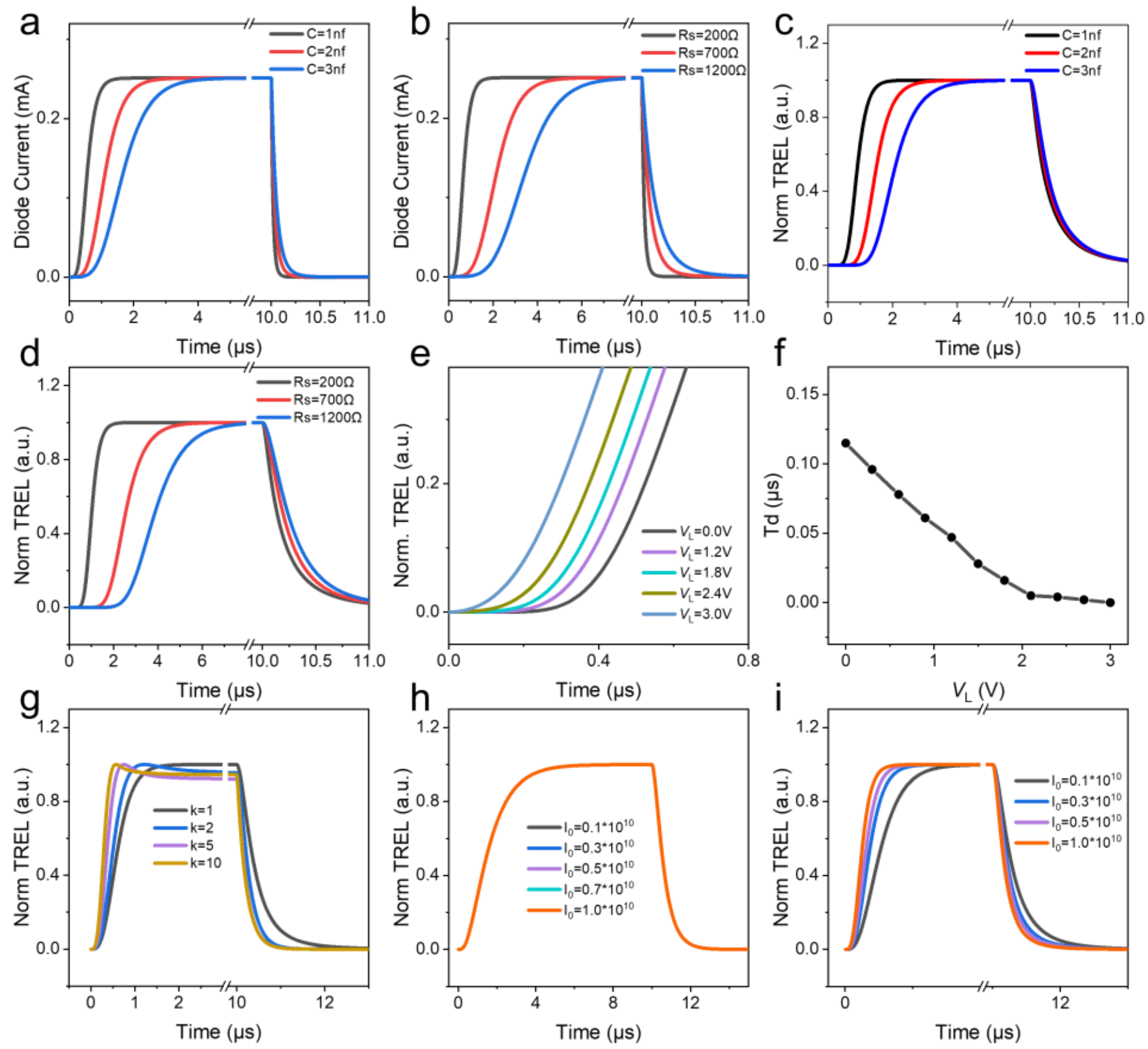

**Figure 3.** Numerical simulation results based on the diode-based RC equivalent circuit model and TREL differential equations in Fig. 2. **(a, b)** Simulated diode current under different capacitances and series resistances. **(c, d)** Simulated normalized TREL curves under different capacitances and series resistances. **(e)** Simulated delay phase characteristics of TREL curves with varying $V_L$ voltage. **(f)** Simulated relationship between TREL curve delay time and $V_L$ voltage. **g)** TREL curves under different injection imbalance coefficients (k). **h)** Normalized TREL curves with only first-order recombination processes at various saturation current densities. **i)** Normalized TREL curves incorporating second-order recombination processes at different saturation current densities.

In the following, we use numerical simulations to discuss the relationship between carrier dynamics and TREL curves, including the balance of electron and hole injection, the features of first- and second-order recombination, the quantitative relationships between TREL response time and the dynamic parameters of TRC and TREL. Mathematically, we used the following approximate function to describe the feature of the diode current curve.

$$I_{inj} = \begin{cases} 0, & t < t_0 \\ I_0 * (1 - e^{-\frac{t-t_0}{\tau}}), & t_0 < t \leq t_{end} \\ I_0 * e^{-\frac{t-t_{end}}{\tau}}, & t \geq t_{end} \end{cases} \quad (7)$$

Here, $I_0$ is the saturated carrier injection rate (units: 1/μs), $\tau$ represents the diode current's rise or fall time, $t_0$ is the delay time of diode current, and $t_{\text{end}}$ is the end time of the square-wave voltage signal, set to 10 μs in our simulation. Specifically, $I_0$ depends on the diode's I-V characteristics, while $\tau$ and $t_0$ are related with charging time and diode turn-on voltage. Because $t_0$ is only related with the time shift of $I_{inj}$ and TREL curves, we set $t_0$ to 0 for simplifying the simulation.

Overshoot is a phenomenon where the luminance initially rises, peaks at a maximum value, and subsequently decays to a steady-state level under voltage pulse application[37-39]. In the following, equation (7) was used to discuss the relationship between injection imbalance and overshoot. A coefficient $k$ of $I_n/I_p$ was introduced to describe injection imbalance, where $I_p$ represents for the saturated carrier injection rate of holes and $k * I_p$ represents for the saturated carrier injection rate of electrons. As shown in Fig. 3g, the imbalance of carrier injection ($k$ > 1) is related with the overshoot of TREL curves, and the overshoot become obviously with $k$ increaseing. As simulated in Figs. S4a-c, the overshoot can be enhanced by increasing $I_p$, or reducing $\tau$, or increasing $k_r$, or their combinations under imbalanced injection. As simulated in Figs. S5a-c, no overshoot can be observed for the devices under balance injection ($k$ = 1) regardless the variation of parameters ($I_p$, $\tau$, $k_r$). Thus, the imbalance injection will lead to accumulation of electron or hole and then lead to overshoot. These results support previous conclusion that overshoot of TREL curves can be correlated with injection imbalance[37,38].

Mathematically, the lifetime in first-order kinetics is concentration-independent, while the second-order kinetics is inversely related to the initial concentration[40]. To qualitatively understand the influence of recombination dynamics on TREL curves, we examined the kinetic characteristics of first and second order recombination processes. If second order recombination is present, both of rising time and decay time decrease with saturation injection current ($I_0$) increasing, as shown in the normalized TREL curves of Fig. 3h. In contrast, normalized TREL curves remain unchanged with the variation of $I_0$ if only first-order recombination exists as shown in Fig. 3i. This result proves the presence of second-order recombination in operated LED devices because both rising and decay time decrease with increasing pulse voltage, as seen in Fig. S6.

Response time of TREL (including rise and fall time) is a critical parameter for optical communication applications[41]. We therefore conducted numerical simulations to quantitatively analyze the correlation between above physical parameters ($I_0$, $k_n$, $k_r$ $\tau$) and response times. Here, we define the rise time ($\tau_{rise}$) as the time for normalized TREL intensity to increase from 0.1 to 0.9, and the fall time ($\tau_{fall}$) as the time for normalized intensity to decrease from 0.9 to 0.1 after voltage Turn-off. For the device under balanced electron-hole injection with equal first-order recombination coefficients ($k_n = k_p$), the response time decrease with $\tau$ decreasing while it decreases with $I_0$, $k_n$, and $k_r$ increasing, as shown in Figs. S7a-d. Furthermore, we derived Equations (S1)-(S5) through combined numerical simulation and theoretical formulation to illustrate the relationships between response time and the combination of $I_0$, $k_n$, $k_r$ $\tau$ (see part 1 of SI for details). Based on the analysis in SI, the response time can be shortened by increasing the injection current density ($I_0$) and the recombination coefficients ($k_r$and $k_n$). Reducing the current time constant ($\tau$) is also effective, as it determines the speed of the current response and sets a fundamental lower bound on the device response time. In practice, τ can be reduced by lowering the RC time constant of the equivalent circuit through device structure optimization, as demonstrated in recent experimental work on high-speed QLEDs[41].

Based on VCE framework, we performed a global fitting analysis by combining TRC and TREL experimental data to extract dynamics parameters for QLED device analysis. As shown in Fig. 4a, a pre-processing method was employed to fit the transient current using peak value ($I_{\mathrm{peak}}$) and balance value ($I_{\mathrm{ss}}$). Firstly, we calculated the equivalent parallel resistance and series resistance under different voltages using $R_{\mathrm{p}}(V) = (V - R_s I_{\mathrm{ss}}(V))/I_{\mathrm{ss}}(V)$ and $R_s$=$V/I_{\mathrm{peak}}$. As shown in Figs. S9a-b, the series resistance is voltage-independent, while the parallel resistance exhibits exponential decrease with increasing voltage. The diode's IV curve was then reconstructed using $R_{\mathrm{p}}(V)$. The $R_s$ and reconstructed IV curve were used to fit the TRC data and extract the capacitance. This pre-processing method offers universal fitting capability for RC parallel circuits with various nonlinear parallel resistances including diode-based RC circuits. Figure 4b shows the fitting results of TRC data for QLED. All TRC fittings achieve $R^2 > 0.99$, confirming the reliability of above fitting. Figures S9b-d displays typical circuit parameters extracted from TRC fitting, including the diode IV curve, series resistance, and capacitance. Using the fitted parameters, we simulated the diode current as a function of time. Figure 4c compares the fitted diode current curve with the measured

TREL curve. The diode current exhibits similar shapes like TREL, which include delay, rise, balance, and decay stages, indicating a correlation between TREL behavior and the diode current. Notably, the fitted delay time of diode current is matched with the measured delay time of TREL curves, although most of the TREL delay times are slightly longer. This phenomenon has been explained in above simulation.

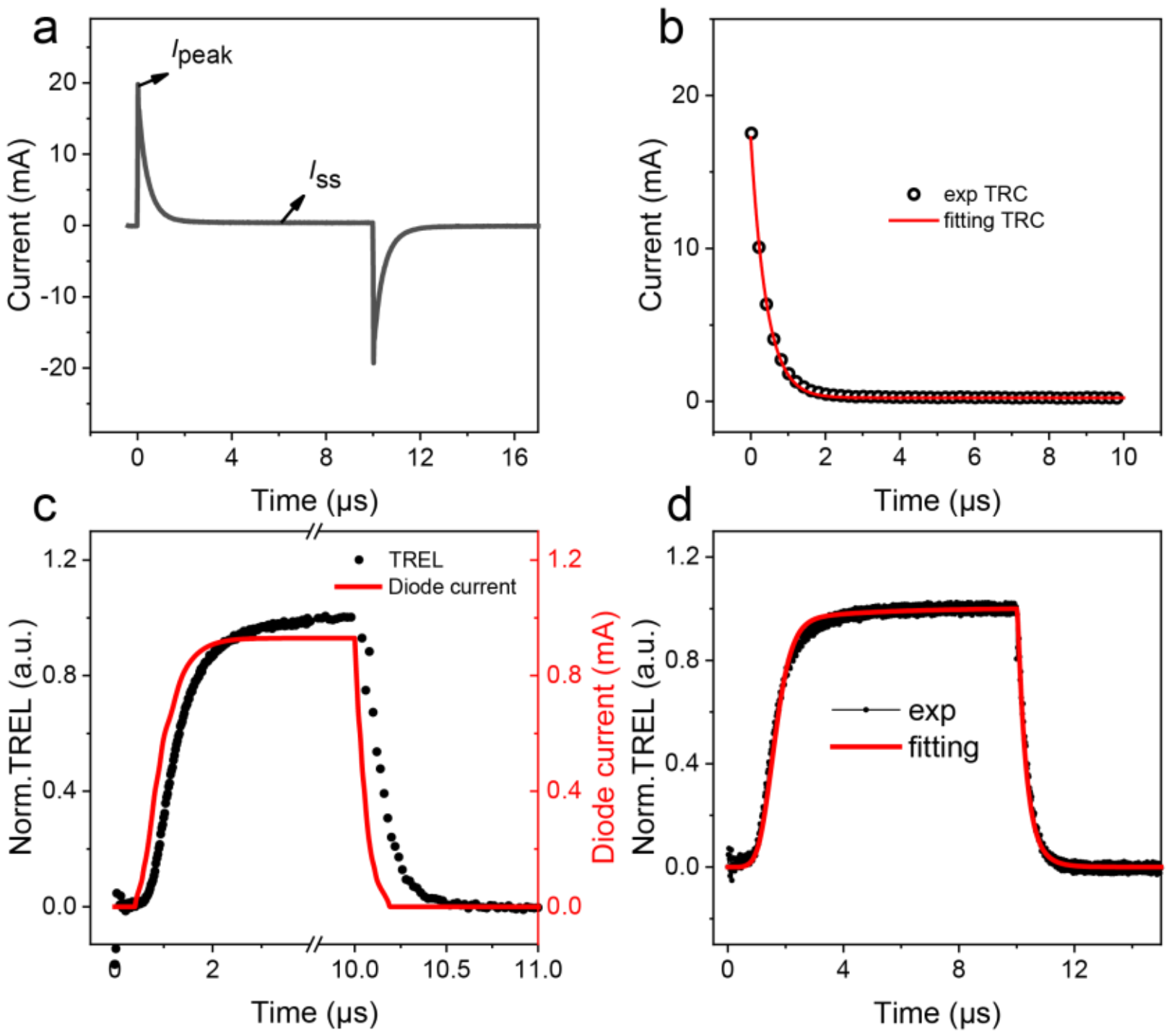

**Figure 4.** a) Transient current (TRC) curve with peak current $I_{\mathrm{peak}}$ and steady-state current $I_{\mathrm{ss}}$ annotation. b) TRC data fitting results using preprocessing method. c) Diode current from TRC fitting and experimental TREL curve comparison. d) Comparison of global fitting results and experimental TREL curves.

Subsequently, the diode current $I_R$ obtained from TRC fitting was converted into carrier injection rates for TREL curve fitting. The electron and hole injection components are defined as $I_{\mathrm{n}}(\mathrm{t}) = A * I_{\mathrm{R}}(\mathrm{t})/q$ and $I_{\mathrm{p}}(\mathrm{t}) = B * I_{\mathrm{R}}(\mathrm{t})/q$, respectively, where $A$ and $B$ represent the ratios of electron and hole injection currents to the total diode current and serve as fitting parameters. Since TREL curves are symmetric with respect to electrons and holes, we set $A > B$ in the fitting process.

Based on the carrier dynamics equations (Equations (5)-(6)), A global fitting was performed to simultaneously describe the delay, rising, steady, and decay stages of the experimental TREL data. Figure 4d and Fig. S10a-d display the fitting curves for typical TREL data, and Figs. S11a-d shows the dynamic parameters extracted from TREL data of a typical QLED using aforementioned global fitting methods. The high coefficients of determination ($R^2 = 0.9971 \pm 0.0015$) support the effectiveness of global fitting and VCE framework in describing the measured TREL and TRC curves. Furthermore, the good consistency indicates that all TREL stages are governed by a single physical mechanism. Beyond this, the global TRC-TREL fitting provides a reasonable and comprehensive extraction of carrier dynamics parameter.

We further tried to fit OLED and GaN-LED using the VCE framework. As shown in Figs. S12–S13. The TREL results of OLED can be well fitted with an average $R^2$ of $0.9906 \pm 0.0016$. For GaN-LED, we considered the barrier capacitance in the equivalent circuit due to their pronounced PN junction features[42]. Equation (8) describes the modified capacitance as the sum of geometric capacitance and barrier capacitance. With these modifications, The TRC and TREL curves of GaN-LED can be well fitted with an average $R^2$ of $0.9995 \pm 0.0002$, as shown in Figs. S14 and S15. These results further demonstrate the applicability of VCE framework in quantitatively describing the carrier dynamics of electroluminescent devices.

$$C(V) = C_g + \frac{C_{j0}}{\sqrt{\left(1 - \frac{V}{V_{bi}}\right)}} \tag{8}$$

To establish a comprehensive understanding of VCE framework, we summarize the characteristic features and key factors of the TREL measurement based on the aforementioned analysis.

For the delay stage, the delaying originates from the initial capacitor charging process. During this process, the injection current of diode is nearly zero. EL signals cannot be observed until the voltage on the diode reach to a threshold. The delay time is mainly related with the diode turn-on voltage, equivalent circuit charging time, and normalization processing, while turn-on voltage is mainly related with the diode characteristics of the device and charging time are correlated with dielectric constant, conductivity, and thickness of functional layers. Therefore, the mobility of functional layer is not directly related with the delay time of LED, but it can affect the charging time and the LED's turn-on voltage of LED due to its strong relationship with conductivity. The correlations between mobility and delay time are complex to be solved. The above conclusions are conflict with calculation of mobility using the delay time of TREL measurements[30,31,32]. For the rising stage, the rising process is correlated with carrier recombination dynamics and the features of transient injection current. For the overshoot in rising stage, numerical simulation illustrates its correlation with electron-hole injection imbalance. For the decay stage, diode current continues to discharge the capacitor at the earlier stage of decay stage. Because of the second order recombination, the decay time decreases with increasing current of balance stage. Therefore, the decay stage of TREL is mainly determined by the decay of diode current and the carrier recombination dynamics.

Based on the principle that voltage drives current, current drives emission of LED operation, we develop the VCE framework for TREL and TRC analysis by combining the diode-based RC equivalent circuit model and the carrier dynamics of electroluminescence processes. Furthermore, it enables global fitting of TREL curves, where a single set of physical parameters can quantitatively fit the complete evolution of the delay, rising, steady, and decay stages with a coefficient of determination exceeding 0.99 across QLED, OLED, and GaN-LED devices. This framework can not only explain the abnormal experimental phenomena that contradict conventional understanding, but also is universally applicable across other LEDs with different materials and device structures. Importantly, this work provides a fundamental physical tool for understanding electroluminescence transient behaviors in LEDs. It offers design guideline for developing high speed LEDs for AI technology, especially optical communication, optical computing.

### Materials and Methods

Experimental details concerning the device/material fabrication, TREL characterization, and data processing are provided in the Supporting Information.

## Acknowledgements

This work is granted by the National Natural Science Foundation of China (52525309 and U23A20683). We thank the MIIT Key Laboratory for Low-Dimensional Quantum Structure and Devices, Experimental Center of Advanced Materials of Beijing Institute of Technology, for technical assistance.

*Correspondence to Haizheng Zhong

E-mail: hzzhong@bit.edu.cn

## Conflict of interest

The authors declare no competing interests.

## Contributions

H.Z. conceived the project and provided guidance. H.B. conceived the VCE framework, performed the experiments, analyzed the data, and developed the numerical fitting code. Y.W. fabricated the OLED devices and conducted the JVL measurements. Y.G. and J.Y. fabricated the GaN-LED devices and conducted the JVL measurements. H.B. and H.Z. wrote the manuscript with input from all authors.

# Support Information for Voltage-Current-Emission Framework of Time-Resolved Electroluminescence

Hui Bao[1], Longjia Wu[2], Yanan Guo[3], Jianchang Yan[4], Ying Wang[5], Haizheng Zhong[1]*

[1]MIIT Key Laboratory for Low-Dimensional Quantum Structure and Devices, School of Materials Science & Engineering, Beijing Institute of Technology, Beijing 100081, China.

[2]New Display Technology Department, TCL Corporate Research,Guangzhou, Guangdong, 510000, China.

[3]School of Integrated Circuit Science and Engineering, Beihang University, Haidian District, Beijing 100191, China.

[4]Research and Development Center for Wide Bandgap Semiconductors, Institute of Semiconductors, Chinese Academy of Sciences, Beijing 100083, China.

[5]Key Laboratory of Photochemical Conversion and Optoelectronic Materials and CityU−CAS Joint Laboratory of Functional Materials and Devices, Technical Institute of Physics and Chemistry, Chinese Academy of Sciences, Beijing, China.

*Corresponding Author's E-mail: hzzhong@bit.edu.cn

## Materials and methods

### Material and Device fabrication

QLEDs were fabricated on ITO glass substrates with a sheet resistance of ∽20 Ω $sq^{-1}$. The substrates were cleaned with deionized water, acetone and isopropanol, consecutively, for 15 min each, and then treated for 15 min with ozone generated by ultraviolet light in air. These substrates were spin-coated with PEDOT:PSS (AI 4803) and baked at 150 °C for 15 min in air. The coated substrates were then transferred to a $N^2$-filled glove box for spin-coating of the TFB, CdSe-ZnS QD and ZnO nanoparticle layers. The TFB layer was spin-coated using 1.5 wt% in chlorobenzene (2,000 rpm. for 30 s), followed by baking at 110 °C for 30 min. This was followed by spin-coating of CdSe–ZnS QDs (20 mg ml-1, toluene) and ZnO nanoparticles (30 mg $ml^{-1}$, ethanol) layers followed by baking at 145 °C for 30 min. The spin concentration of TFB varied from 4 mg $mL^{-1}$ to 16 mg $mL^{-1}$ to control TFB thickness. The spin concentration of QDs varied from 10 mg $mL^{-1}$ to 30 mg $mL^{-1}$ to control QD thickness. The spin concentration of ZnO varied from 15 mg $mL^{-1}$ to 50 mg $mL^{-1}$ to control ZnO thickness. These multilayer samples were then loaded into a custom high vacuum deposition chamber (background pressure, ∽3 × $10^{-7}$ torr) to deposit the top Al cathode (100 nm thick) patterned by an in situ shadow mask to form an active device area of 4 $mm^2$.

For GaN-LED, the blue LED wafer was grown on c-plane sapphire substrate by metal-organic chemical vapor deposition (MOCVD). The epilayer structure comprises a 2-μm-thick undoped GaN template, a 1-μm-thick Si-doped n-type GaN layer, 10 periods of $In_{0.12}Ga_{0.88}N$/GaN (3 nm/10 nm) multiple quantum wells (MQWs), a 30-nm-thick Mg-doped $Al_{0.15}Ga_{0.85}N$ electron blocking layer and a 150-nm-thick Mg-doped p-GaN contact layer. After the epitaxy, standard chip fabrication process in a face-up configuration was performed. Indium tin oxide (ITO) was deposited on the wafer to serve as the transparent conductive layer. The chip mesa structures were then defined via inductively coupled plasma (ICP) etching to expose the n-GaN layer. Subsequently, Cr/Pt/Au metal stacks were deposited by e-beam evaporation to form the n-type and p-type electrodes, respectively. The wafer was then back grinded and thinned to a thickness of 150 μm, and individual chips with an area of 45 × 45 $mil^2$ were separated using the nanosecond (ns) pulse laser scribing. The LED chips were further mounted onto lead frames and wire-bonded with gold wires. The electroluminescence (EL) spectra and light output power (LOP) of the unencapsulated LEDs were measured at room temperature in a calibrated integrating-sphere system equipped with an EVERFINE® HAAS-2000 high-accuracy array spectroradiometer.

OLEDs were fabricated on glass substrates coated with indium tin oxide (ITO) having a sheet resistance of approximately 20 Ω $sq^{-1}$. The substrates were cleaned sequentially with deionized water, acetone, and isopropanol for 15 minutes each under ultrasonic agitation. Subsequently, the substrates were treated with ozone generated by ultraviolet light in air for 15 minutes to enhance surface hydrophilicity. After cleaning, the substrates were immediately transferred into a high-vacuum deposition chamber (base pressure ~$3 \times 10^{-7}$ torr) without exposure to air, and all organic layers as well as the aluminum cathode were thermally evaporated in the same vacuum run, with thicknesses monitored in situ by a quartz crystal microbalance. The layers were deposited in the following order. A 10nm-thick layer of HAT-CN (hexaazatriphenylenehexacarbonitrile), a 160 nm layer of TAPC (1,1-bis[(di-4-tolylamino)phenyl]cyclohexane), a 10nm layer of TCTA (4,4′,4′′-tris(carbazol-9-yl)-triphenylamine), a 20 nm co-evaporated emissive layer of CBP (4,4′-bis(N-carbazolyl)-1,1′-biphenyl) doped with 4 wt % of an iridium complex, a 35 nm layer of TmPyPB (1,3,5-tri(m-pyrid-3-yl-phenyl)benzene), a 2 nm layer of Liq, and a 100 nm aluminum (Al) cathode were deposited sequentially.

## Characterizations

TREL and TRC measurements were performed using a home-built setup consisting of a digital oscilloscope (DPO 7104, Tektronix, Beaverton, OR, USA), a photodetector (HAMAMATSU 012702-11, Hamamatsu Photonics, Hamamatsu, Japan), a voltage pulse generator (SDG 5162, Siglent, Shenzhen, China), a variable resistor, and BNC coaxial cables.

LED device characterization was performed at ambient temperature utilizing a J-V-L (current density-voltage-luminance) measurement system comprising: a photodetector (Thorlabs FDS1010), fiber-optic spectrometer (Avantes HSC-TEC), 6-inch integrating sphere, and source measurement unit (Keithley 2400). External quantum efficiency (EQE) values were calculated through the Lambertian emission model following established methods. Fig. S16 displays the IVL characteristics of one representative device from each of the three LED types (GaN-LED, OLED, and QLED) used in this work.

## Data process methods

In this paper, the data processing workflow of the global TREL fitting code is as follows. 1) Time, TRC, and TREL signals are extracted from CSV files at each bias voltage. The series resistance is calculated by extrapolating the current peak, and an interpolation function of the diode I-V curve is reconstructed from steady-state current-voltage data. This data preprocessing method can be applied to various types of nonlinear parallel resistance, not just diodes. 2) For each voltage, a fitting is carried out using an RLC transient model. The inductance in the model comes from the wires of the measurement system, while the LED device itself can still be treated as an RC circuit. A differential evolution algorithm is used to simultaneously optimize the capacitance-related parameters and the wire inductance, yielding an accurate diode current waveform. 3) The fitted diode current serves as the carrier injection term and is fed into a rate equation model for a two-stage fitting of the normalized TREL decay curve. Differential evolution first performs a global optimization, and then the trust-region reflective method refines the result and provides the covariance matrix, from which confidence intervals and correlations are computed. During the TREL fitting process, we impose a constraint to keep the carrier injection rate essentially non-decreasing as the bias voltage increases. All fitting results and intermediate variables are finally summarized and exported to Excel and image files.

## Part 1: response time analysis

At first, we find that the rise time and fall time of the TREL curve exhibit the following relationship with $I_0$, $k_n$, and $k_r$ under pulse current drive.

$$\tau_{rise} \propto \frac{1}{\sqrt{I_0 * k_r + \left(\frac{k_n}{2}\right)^2}} \quad \text{S1}$$

$$\tau_{fall} \propto \frac{1}{k_n} * ln \frac{\frac{k_n}{0.316} + \sqrt{I_0 * k_r}}{\frac{k_n}{0.948} + \sqrt{I_0 * k_r}} \quad \text{S2}$$

When the injection current follows the approximate form given in Equation (7), characterized by a rise/fall time constant τ, the response time is not only bounded by τ but also depends on the interplay between first- and second-order recombination. Through a number of numerical simulations, we derive the empirically modified expressions, defined as follows.

$$T_1 = \frac{1}{\sqrt{I_0 * k_r + \left(\frac{k_n}{3.6}\right)^2}} \quad \text{S3}$$

$$T_2 = \frac{1}{k_n} * ln \frac{\frac{k_n}{1.8} + \sqrt{I_0 * k_r}}{\frac{k_n}{0.6} + \sqrt{I_0 * k_r}} \quad \text{S4}$$

Then we simulated how variations in $I_0$, $k_n$, $k_r$ affect the response time of the TREL curve at different $\tau$ and generated the corresponding $\tau_{\text{rise}}$-$T_1$ and $\tau_{\text{fall}}$-$T_2$ plots. As shown in Fig. S8, we fitted the functional relationships between $\tau_{\text{rise}}$ and $T_1$ as well as between $\tau_{\text{fall}}$ and $T_2$ using Equation S5.

$$y = ax + b + c * e^{-dx} \quad \text{S5}$$

Here a, b, c, and d are fitting coefficients, where x is $\tau_{\text{rise}}$ (or $\tau_{\text{fall}}$) and y is the corresponding rise time (or fall time). Using Equation S5, we fitted the curves in Fig. S8. Equation (S5) provides a good description of the simulated data across the entire range of τ values.
The parameter ranges used in the numerical simulations are as follows. $0.1 < \tau < 1.0$ (μs), $1 \times 10^9 < I_0 < 1 \times 10^{10}$ $(\frac{1}{\mu s})$, $1 \times 10^{-10} < k_r < 1 \times 10^{-8}$ $(\frac{1}{\mu s})$, $0.1 < k_n < 1$ $(\frac{1}{\mu s})$. The simulation parameter ranges were determined based on experimentally fitted values. This empirical formula approximates the relationship between physical parameters and response times in the model, serving as a reference framework for related research.
Based on the above analysis, the response time of TREL can be effectively shortened by increasing the injection current density ($I_0$), the bimolecular recombination coefficient ($k_r$), and the first-order loss coefficient ($k_n$), while reducing the current rise/fall time constant (τ). Among these factors, τ sets a fundamental lower bound on the response time—no response can be faster than the current injection itself. The effects of $k_r$ and $k_n$ are distinct: increasing $k_r$ accelerates radiative recombination and thus shortens both rise and fall times, while increasing $k_n/k_p$ accelerates carrier depletion and primarily shortens the fall time, albeit at the cost of reduced efficiency. These results provide clear guidelines for device design: to achieve faster TREL response without sacrificing efficiency, the priority should be to reduce the RC time constant (thereby reducing τ) and to enhance the radiative recombination coefficient ($k_r$), while minimizing trap-assisted recombination ($k_n$) to maintain high efficiency.

It is worth noting that, from a purely kinetic standpoint, increasing $k_n$ alone would also shorten the response time.

**Part 2: figures**

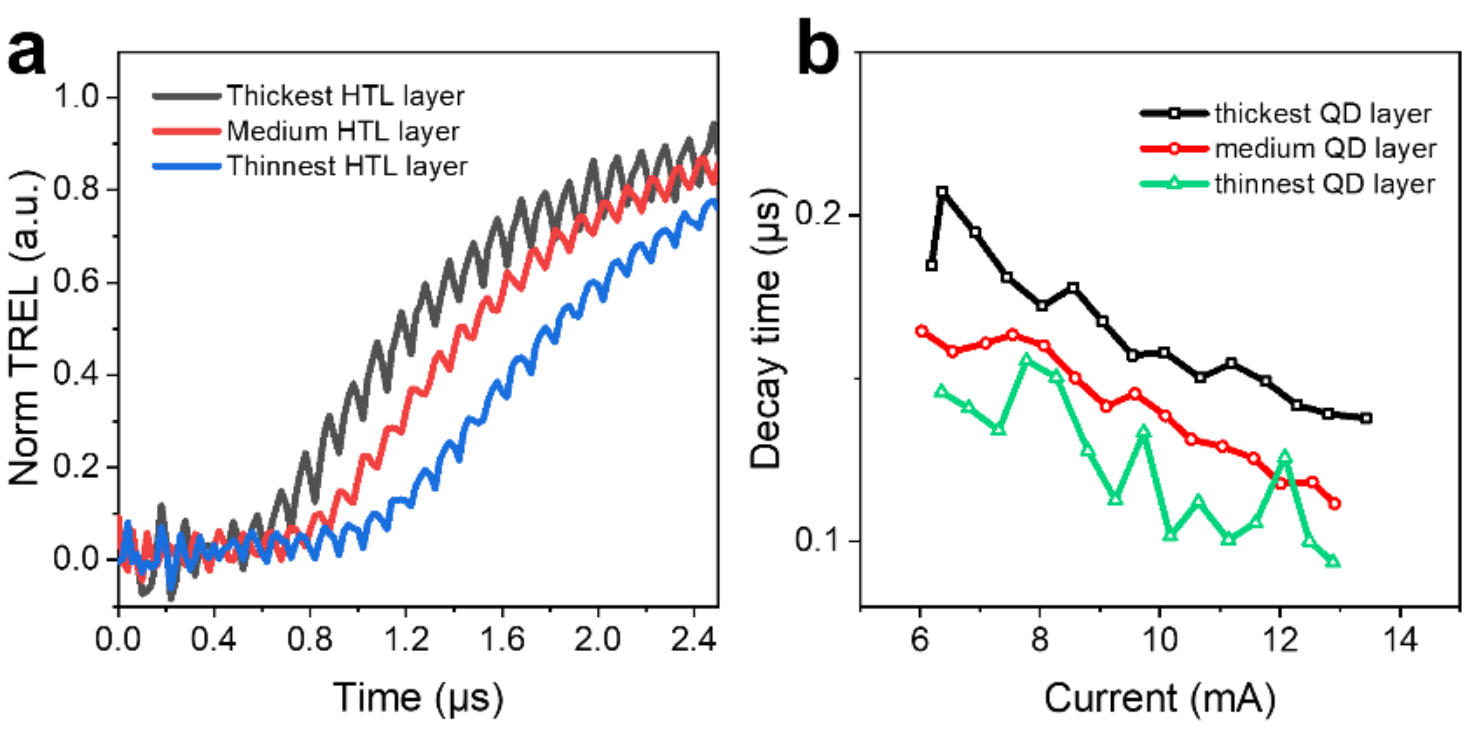

**Figure S1.** (a) Delay stage of TREL curves for QLEDs with different HTL thicknesses. (b) Decay time of TREL curves versus balance current for QLEDs with different QD thicknesses.

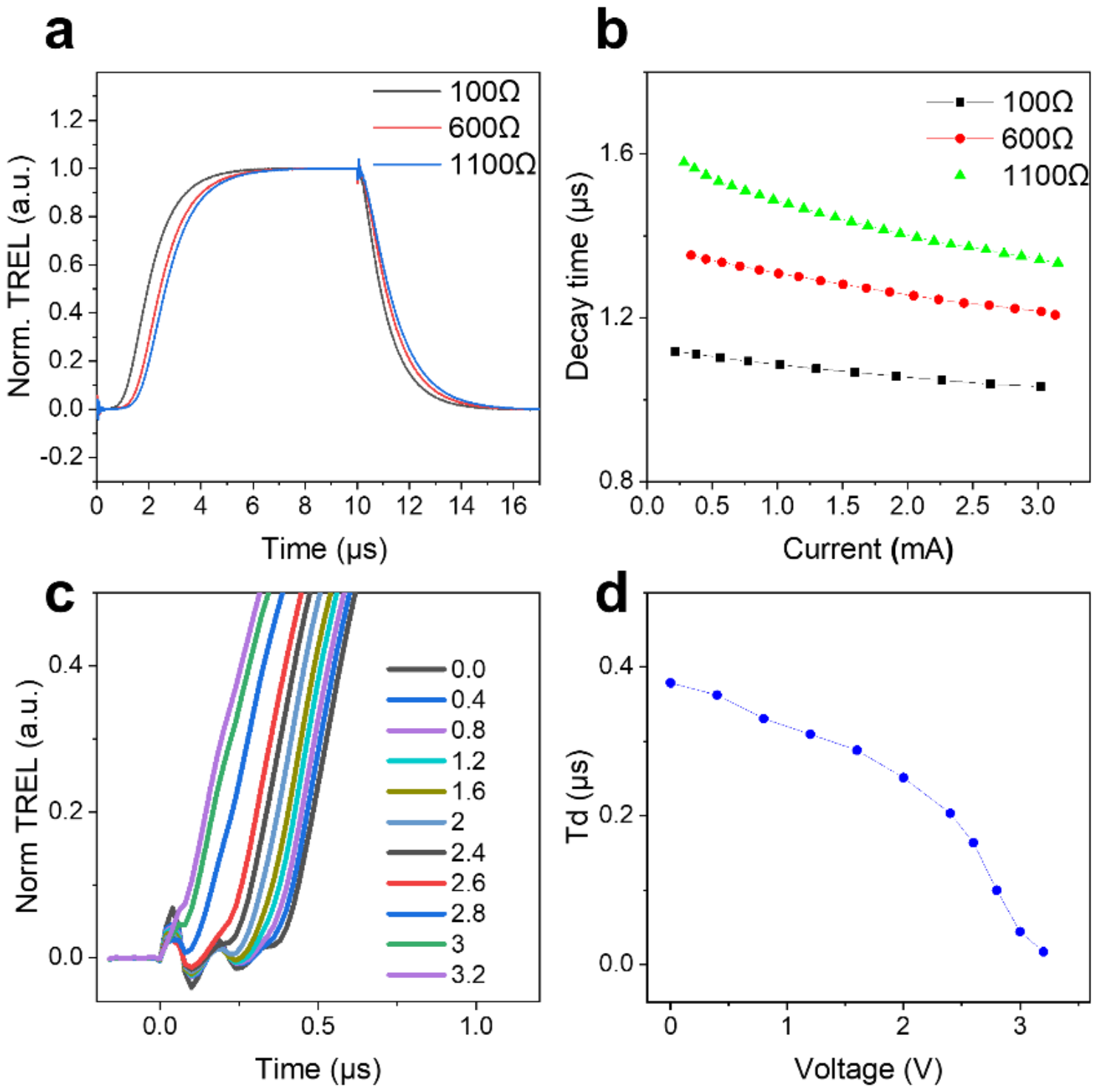

**Figure S2.** (a) TREL curves of OLED devices under different external resistances. (b) Decay time of TREL curves versus balance current for OLEDs with different external resistances. (c) TREL curves in delay stage for OLED devices at varying $V_L$ voltages. (d) Delay time versus $V_L$ for OLED devices.

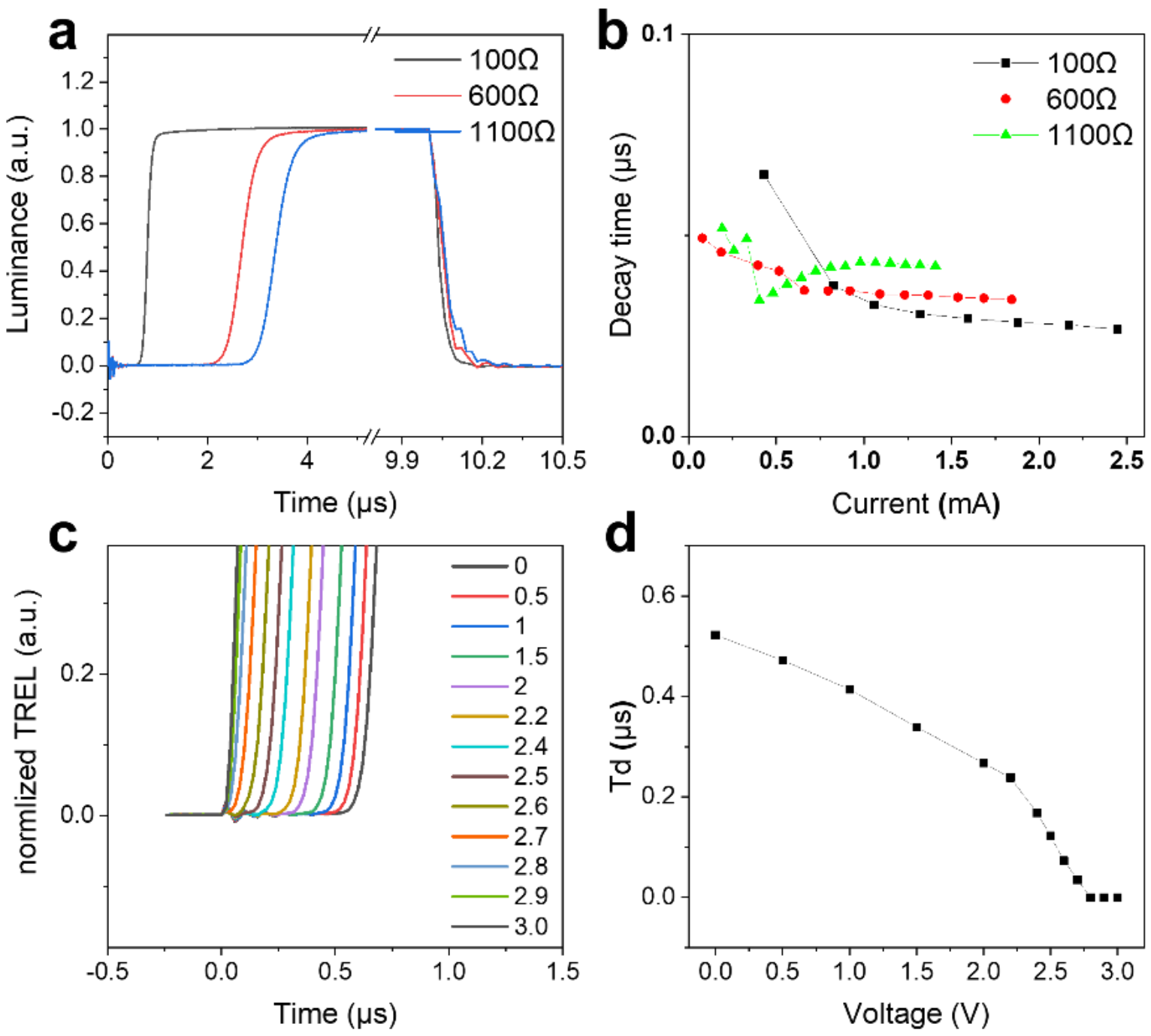

**Figure S3.** (a) TREL curves of GaN LED devices under different external resistances. (b) Decay time of TREL curves versus balance current for GaN LEDs with different external resistances. (c). TREL curves in delay stage for GaN LED at varying $V_{\mathrm{L}}$ voltages. (d). Delay time versus $V_{\mathrm{L}}$ for GaN LED devices.

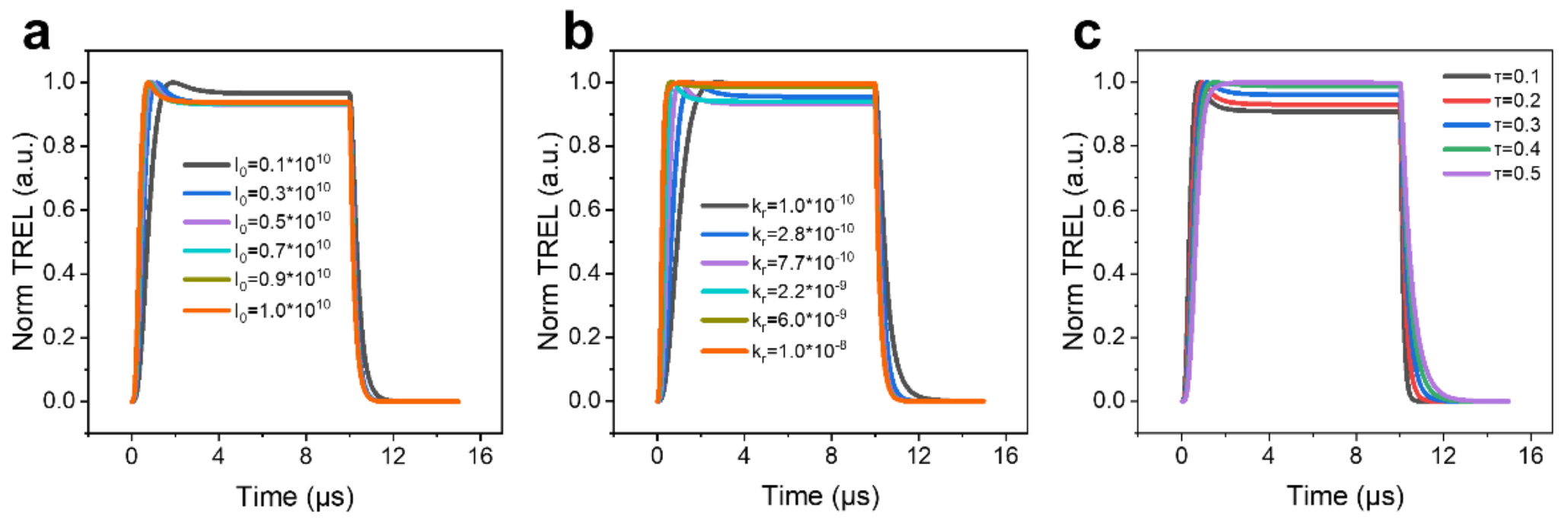

**Figure S4.** Numerical simulated TREL curves at $K$=3 for varying (a) $I_0$, (b) $k_r$, and (c) $\tau$.

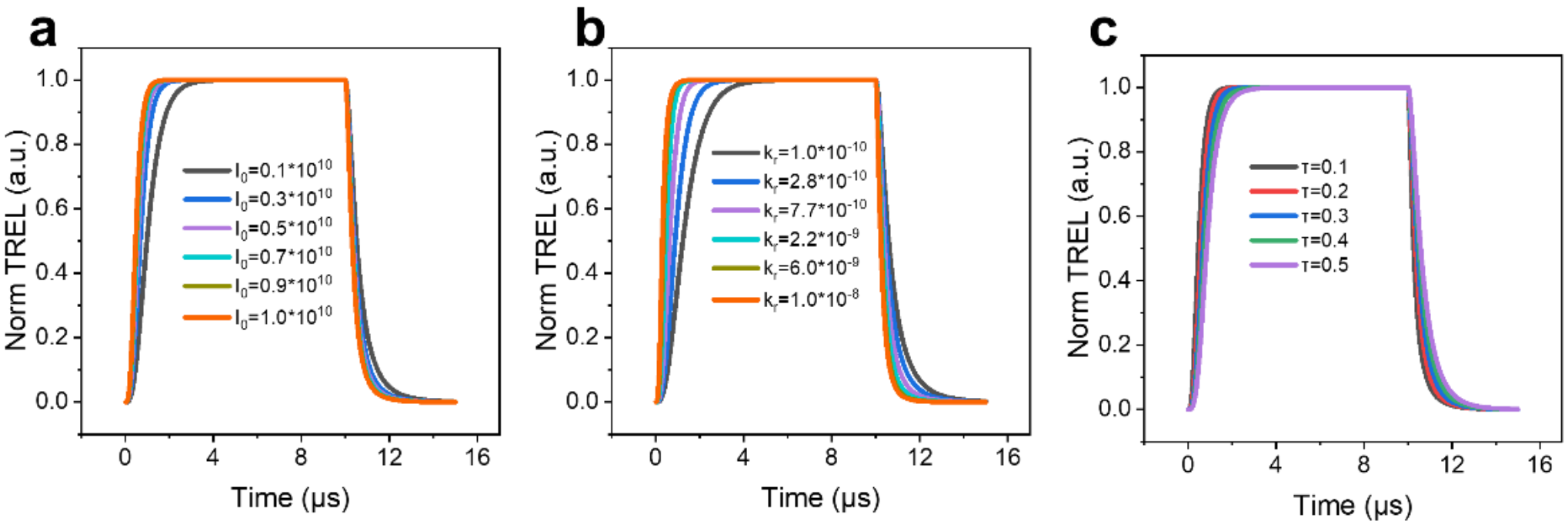

**Figure S5.** Numerical simulated TREL curves at k=1 for varying (a) $I_0$, (b) $k_r$, and (c) $\tau$.

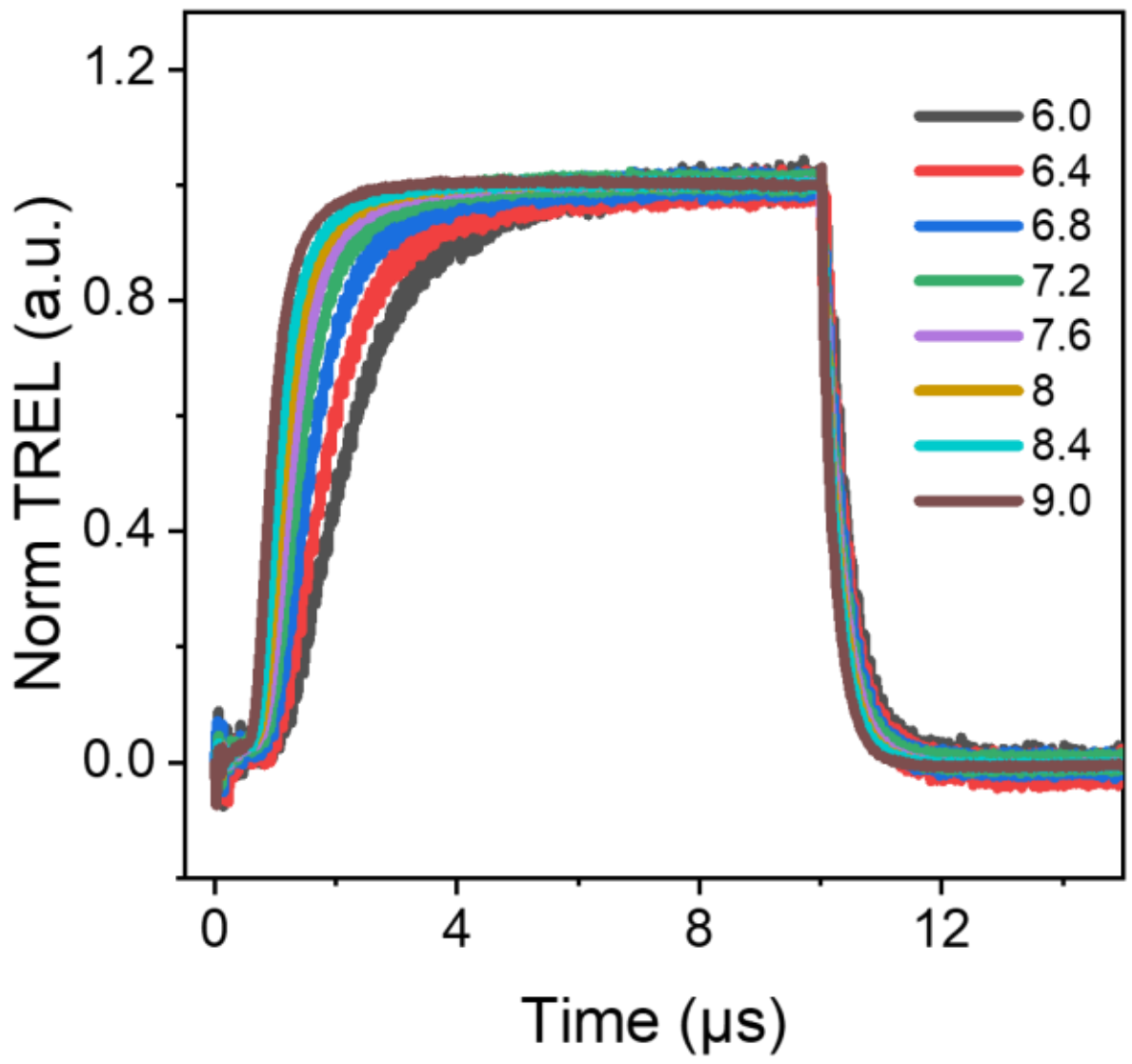

**Figure S6.** TREL measurement curves of QLED devices under different applied voltages.

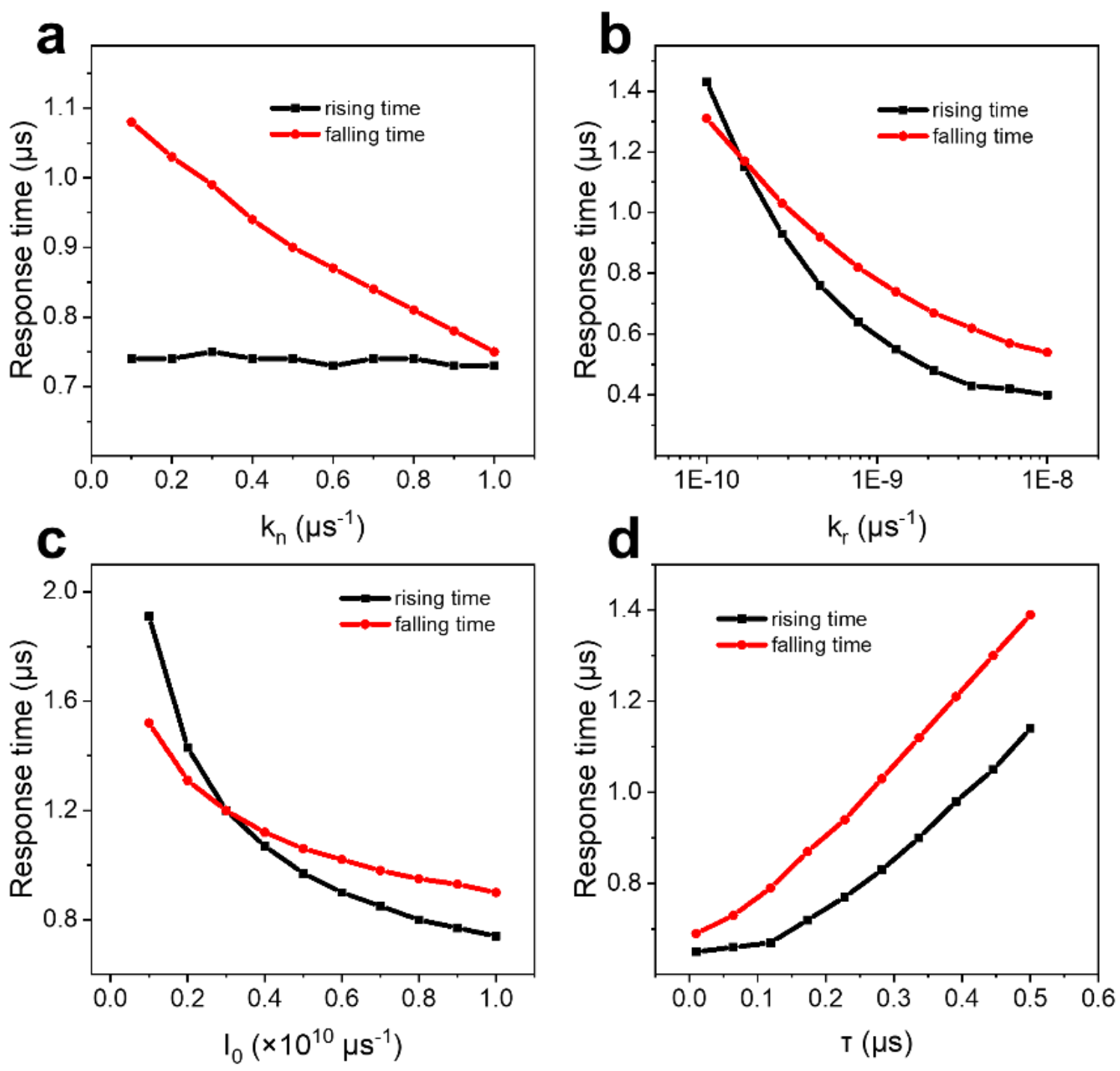

**Figure S7.** Simulated results of response time at varying (a) $K_n$, (b) $k_r$, (c) $I_0$, and (d) $\tau$.

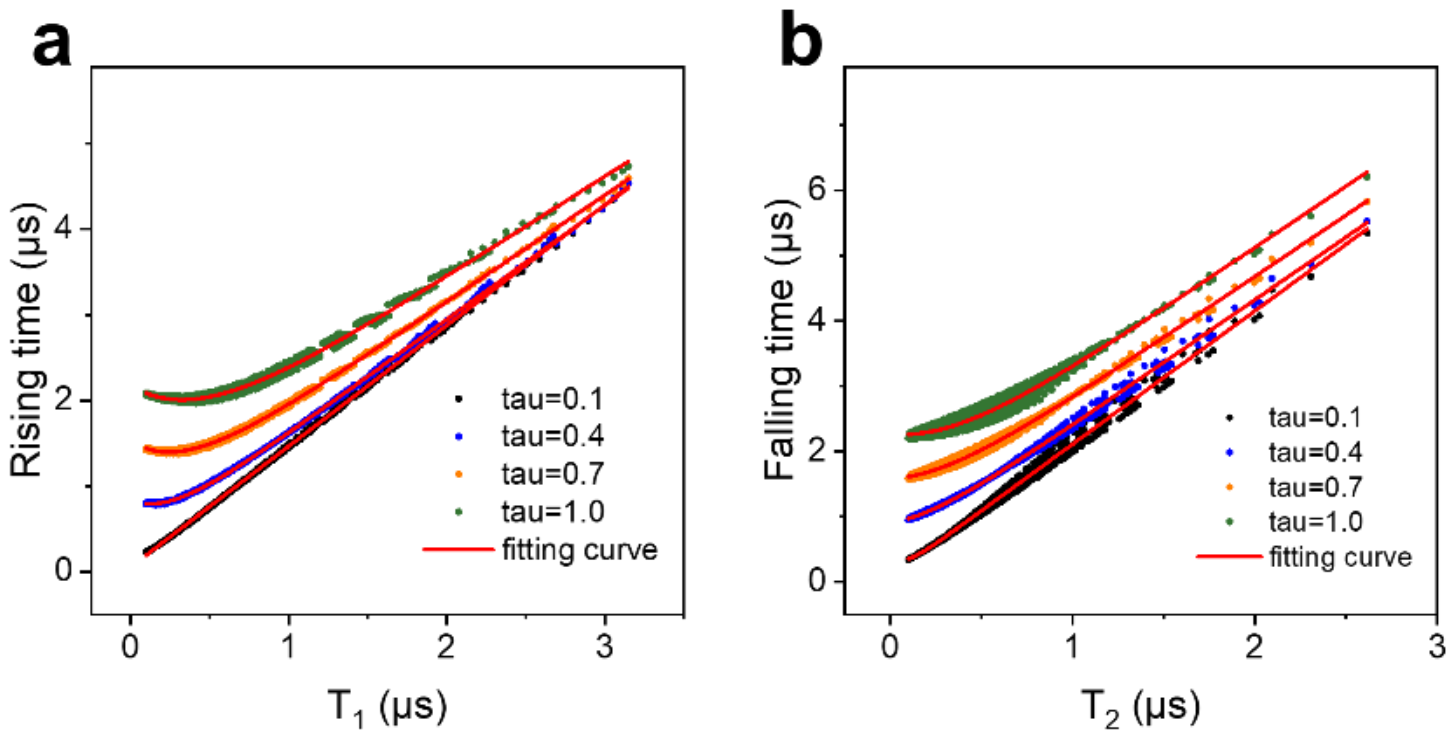

**Figure S8.** Simulated results and fitting curves using Eqs. S3-S5 for TREL (a) rise time and (b) fall time.

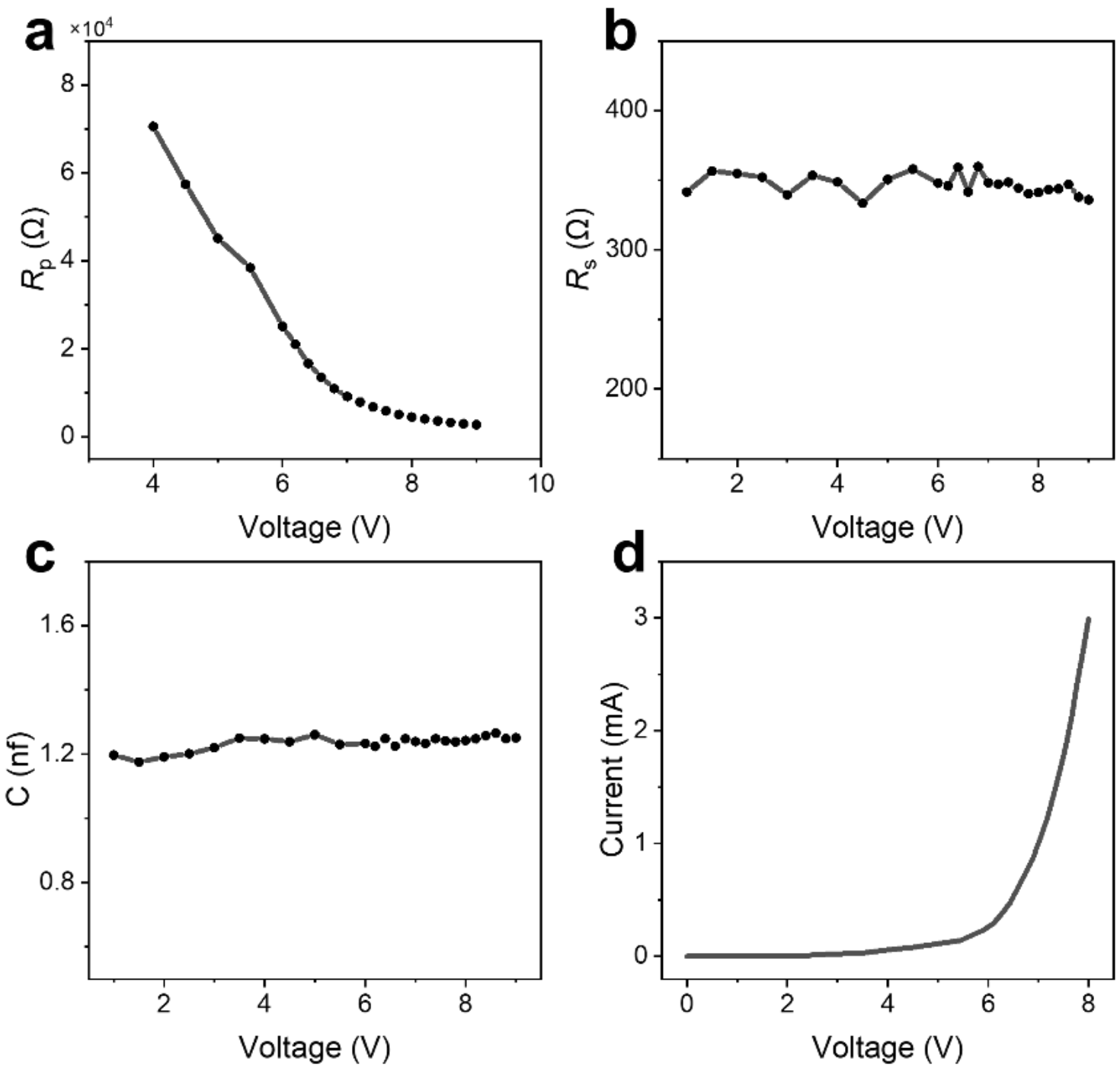

**Figure S9.** (a) $R_p$, (b) $R_s$, and (c) $C$ versus voltage from TRC fitting, with (d) reconstructed diode $I$-$V$ curve using $R_p$.

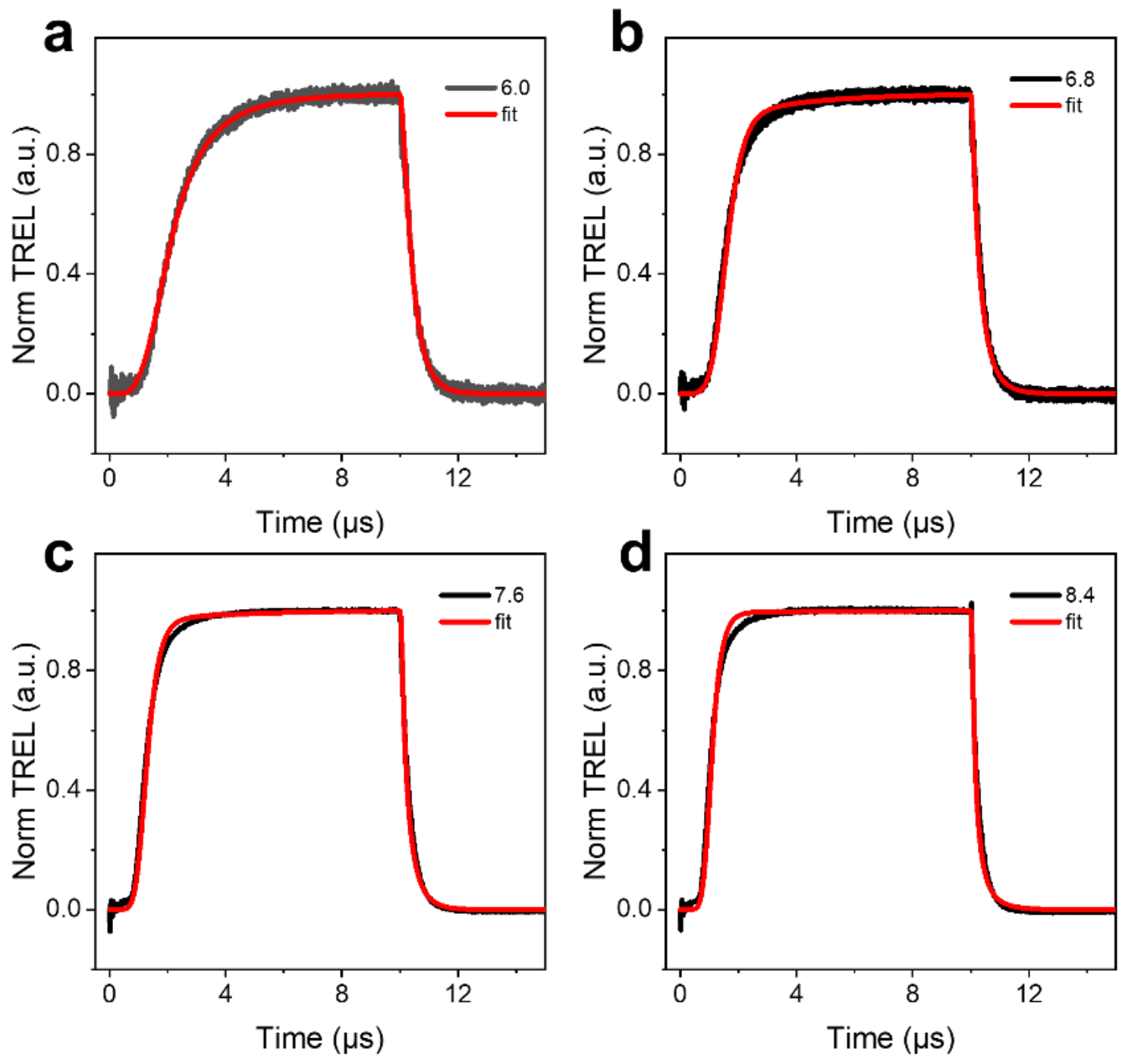

**Figure S10.** Global fitting results of QLEDs' TREL curves at (a) 6.0 V, (b) 6.8 V, (c) 7.6 V, and (d) 8.4 V.

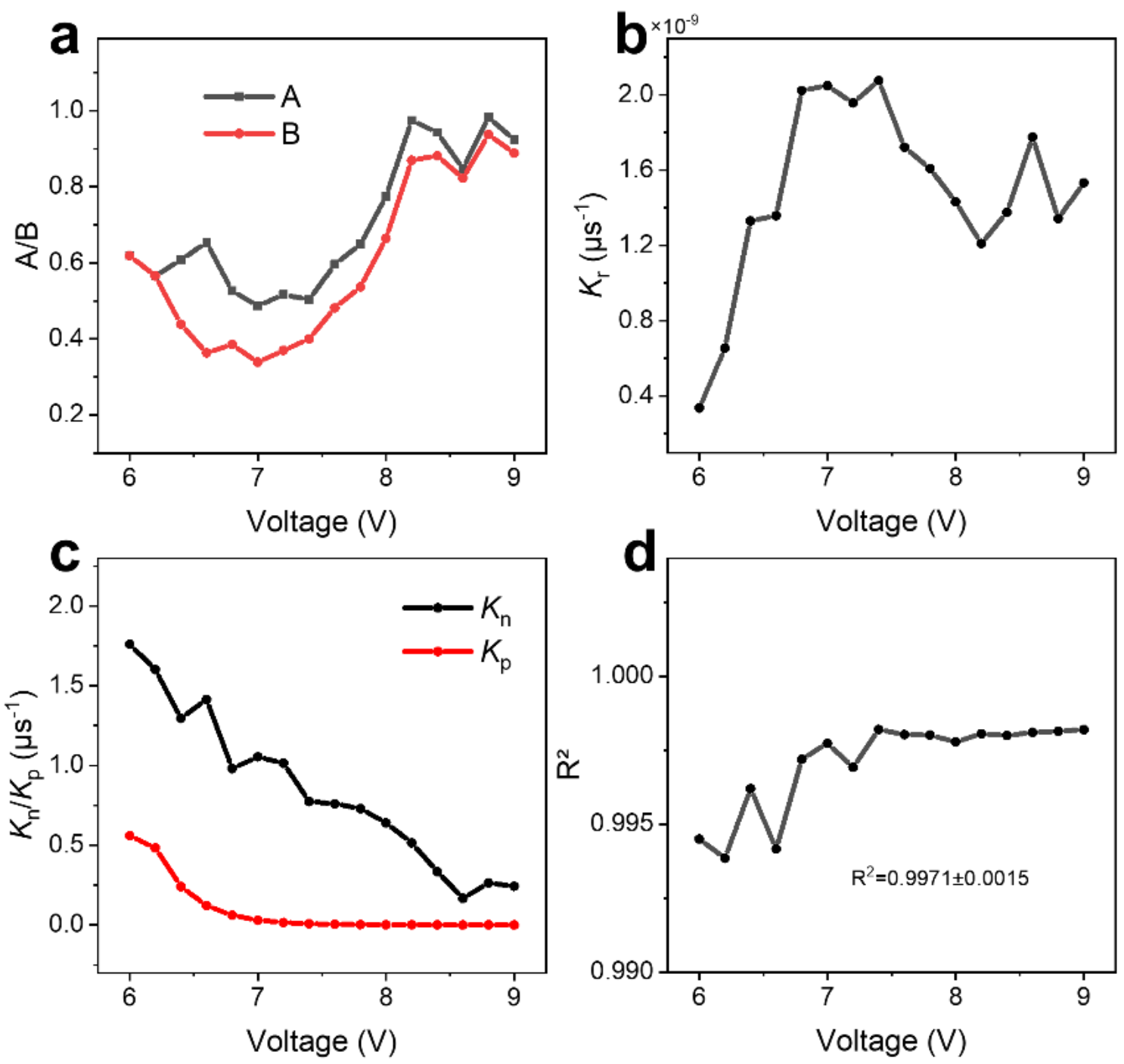

**Figure S11.** Carrier dynamics parameters versus voltage extracted from global fitting of QLEDs' TREL curves. (a) $A/B$, (b) $k_r$, (c) $K_n/K_p$, and (d) coefficient of determination ($R^2$).

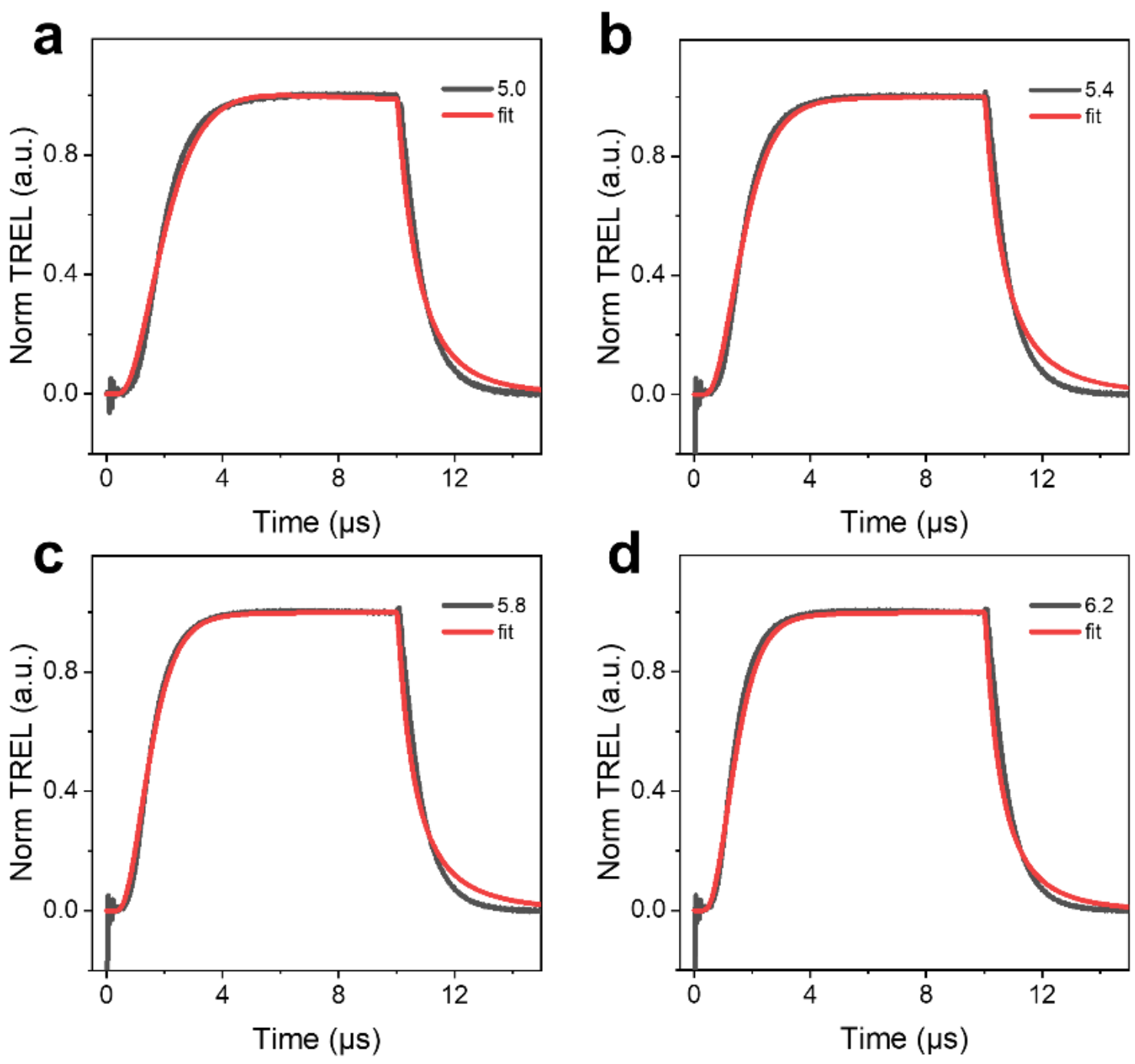

**Figure S12.** Global fitting results of OLEDs' TREL curves at (a) 5.0 V, (b) 5.4 V, (c) 5.8 V, and (d) 6.2 V.

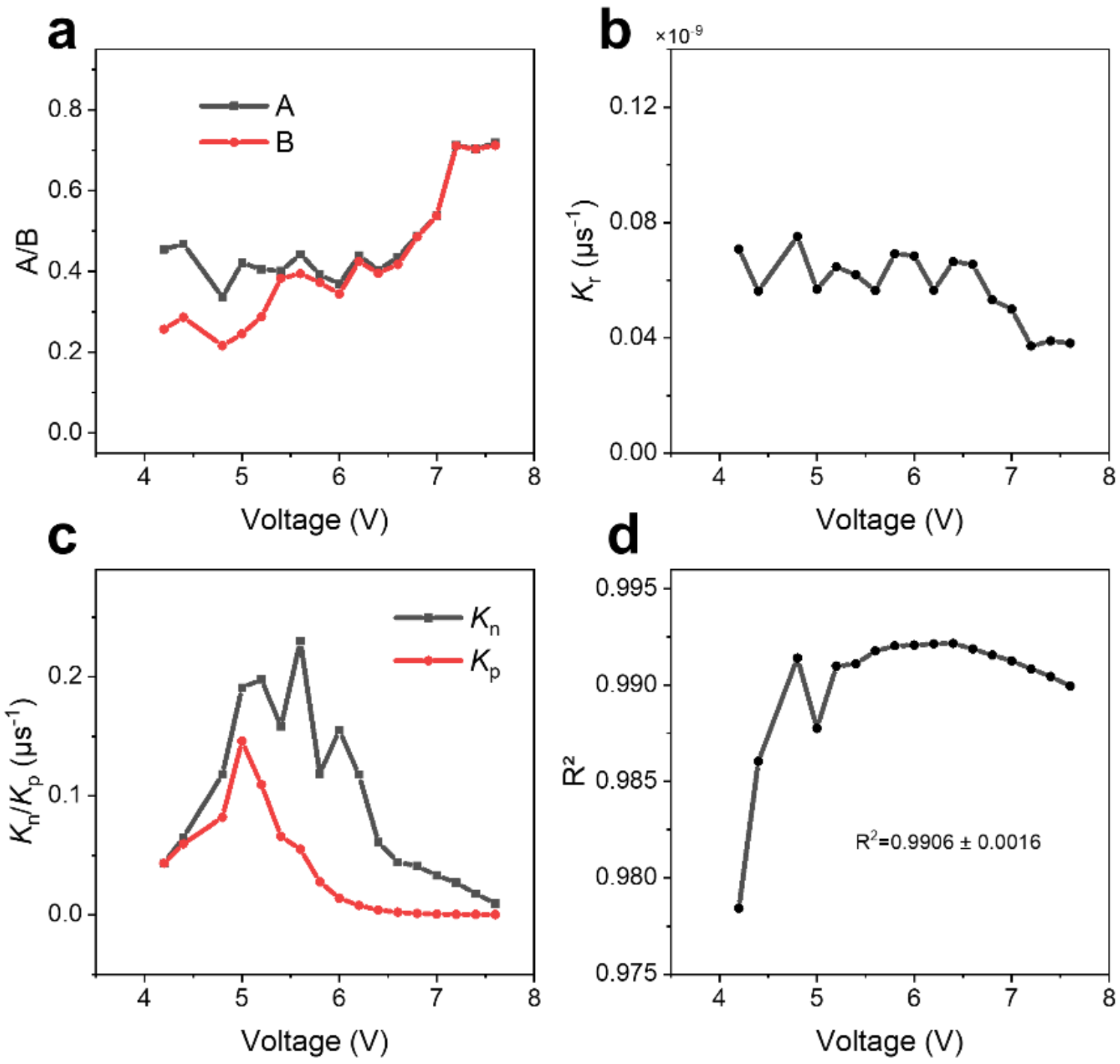

**Figure S13.** Carrier dynamics parameters versus voltage extracted from global fitting of OLEDs' TREL curves. (a) $A/B$, (b) $k_r$, (c) $K_n/K_p$, and (d) coefficient of determination ($R^2$).

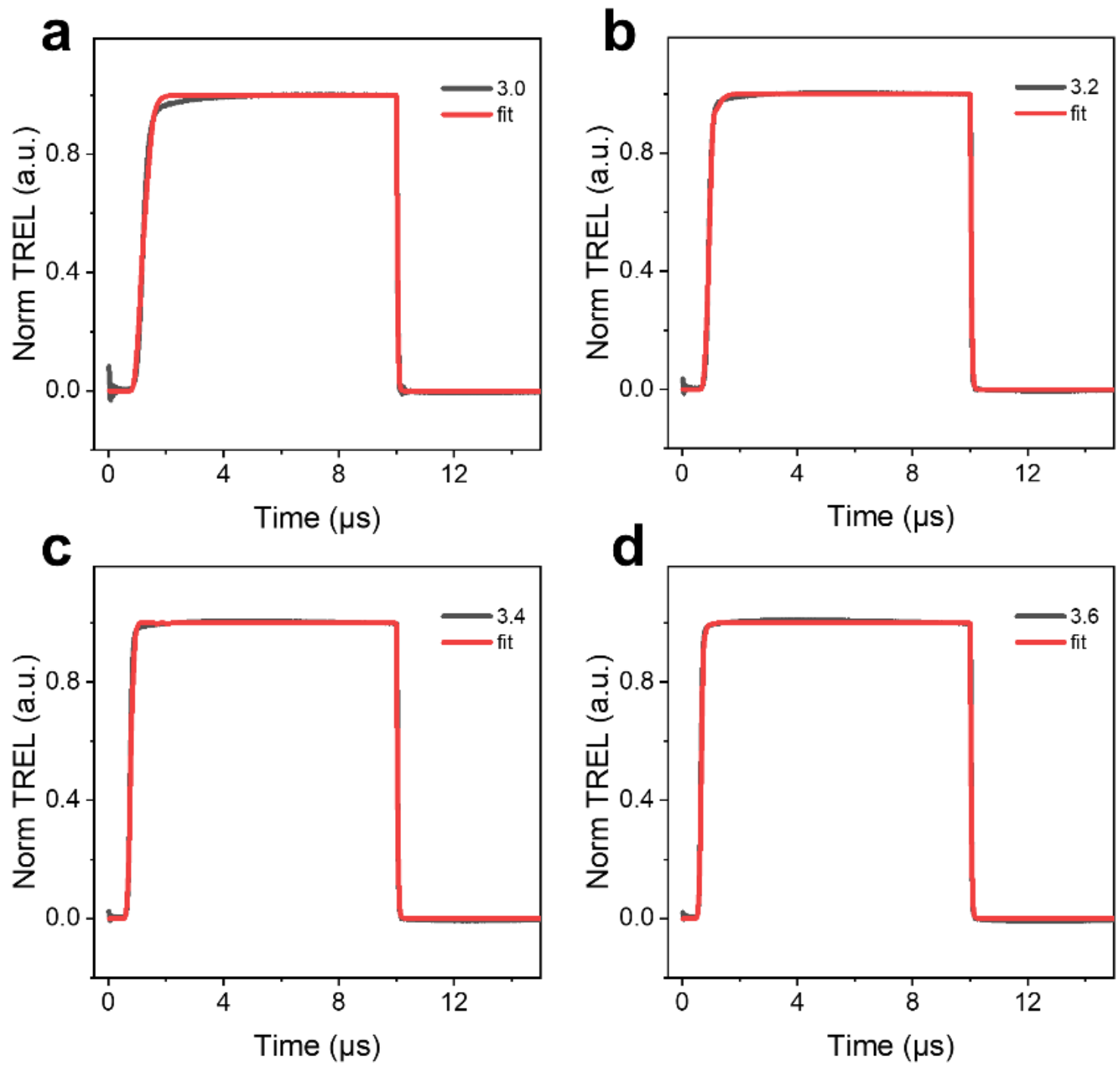

**Figure S14.** Global fitting results of GaN-LEDs' TREL curves at (a) 5.0 V, (b) 5.4 V, (c) 5.8 V, and (d) 6.2 V.

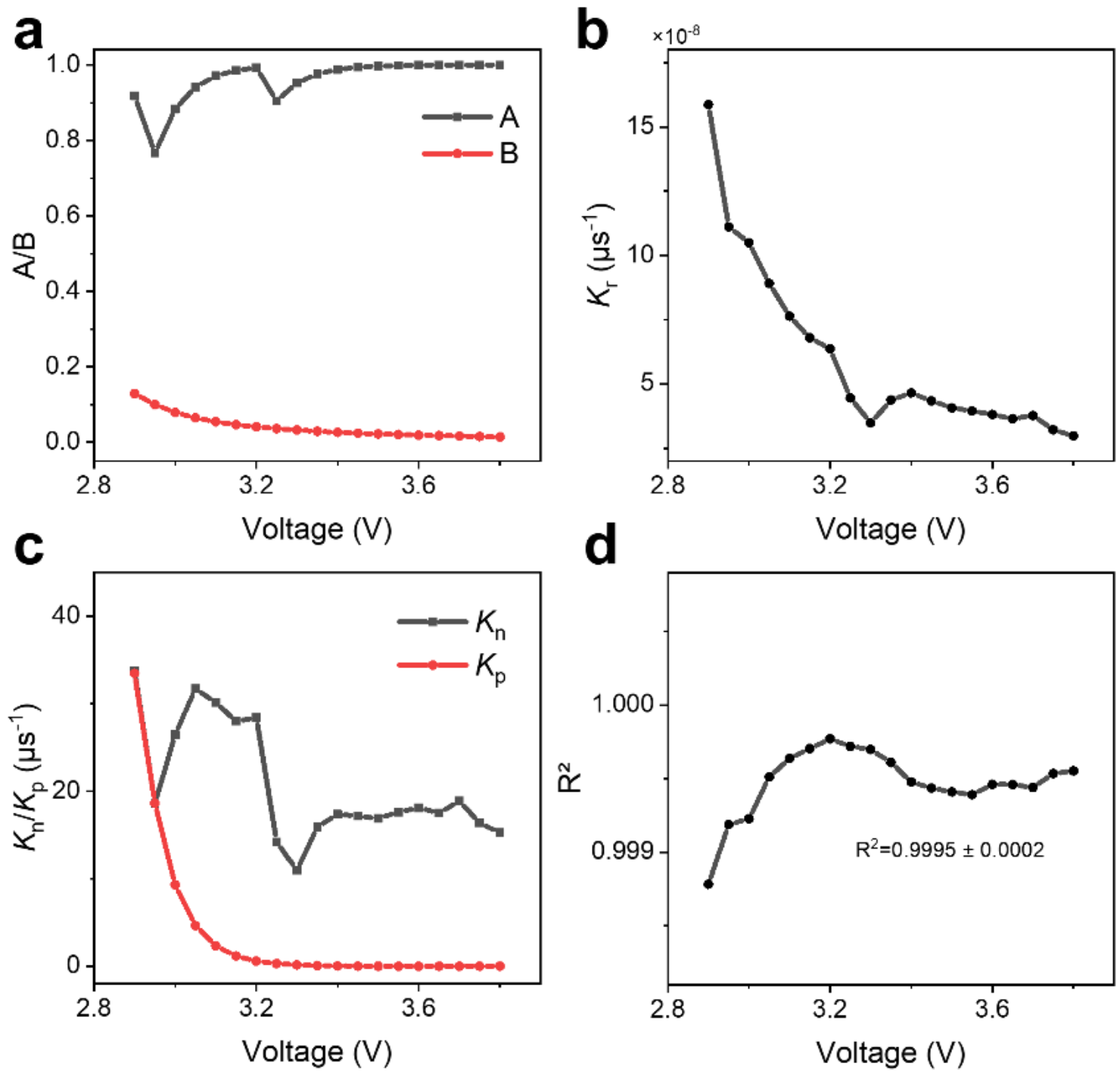

**Figure S15.** Carrier dynamics parameters versus voltage extracted from global fitting of GaN-LEDs' TREL curves. (a) $A/B$, (b) $k_r$, (c) $K_n/K_p$, and (d) coefficient of determination ($R^2$).

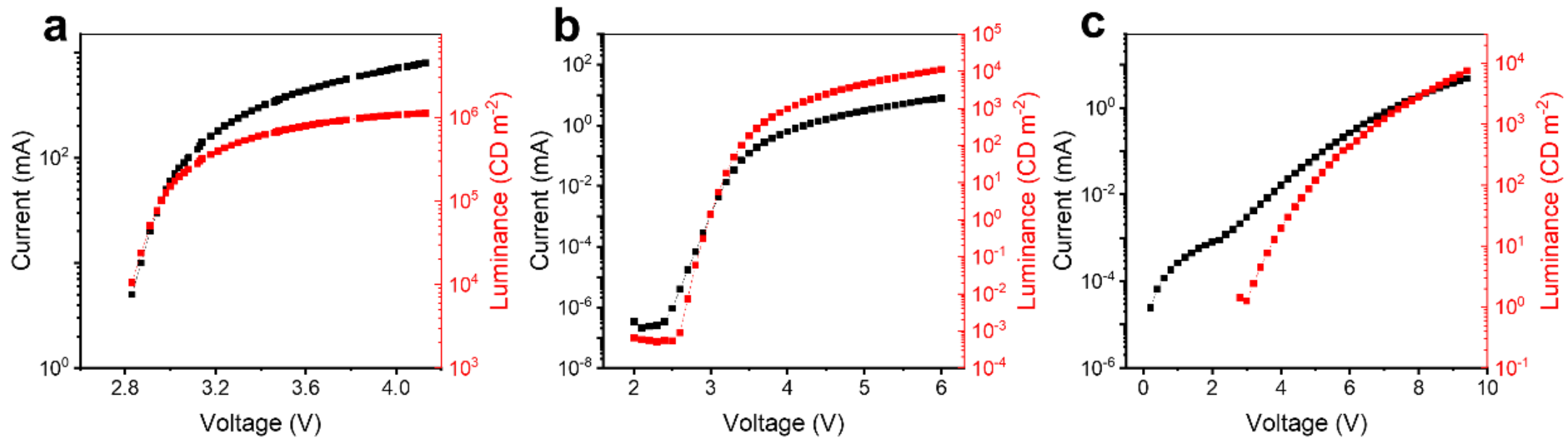

**Figure S16.** IVL curves of a typical (a) QLED, (b) OLED and (c) GaN-LED